\documentclass[12pt]{iopart}
\usepackage{amssymb}
\usepackage{epic}
\usepackage{eepic}
\usepackage{amsfonts}
\usepackage{bm}

\newcommand{\nc}{\newcommand}
\nc{\JStP}{{\it J. Stat. Phys.}}  \nc{\IJMP}{{\it Intern. J. Mod. Phys.}}
\nc{\BBS}{{\rm BBS}\ }
\nc{\g}{\gamma}
\nc{\lm}{\lambda} \nc{\la}{\lambda}
\nc{\bh}{{\bf h}}
\nc{\cR}{{\cal R}} \nc{\kp}{{\varkappa}} \nc{\om}{\omega}   \nc{\CN}{{\cal N}}
\nc{\qt}{\tilde{q}} \nc{\tp}{\tilde{p}} \nc{\rt}{\tilde{r}}
\nc{\ty}{\tilde{y}} \nc{\tx}{\tilde{x}} \nc{\tQ}{{\widetilde Q}}
\nc{\trh}{\tilde{\rho}}
\nc{\ny}{\nonumber}
\nc{\lk}{\left(} \nc{\rk}{\right)} \nc{\Rb}{\right]} \nc{\Lb}{\left[}
\nc{\rb}{\right\}} \nc{\lb}{\left\{}
\nc{\hs}{\hspace*{1cm}} \nc{\hx}{\hspace*{3mm}} \nc{\hq}{\hspace*{6mm}}
\nc{\eR}{{\rm R}} \nc{\eL}{{\rm L}}
\nc{\CD}{{\cal D}}  \nc{\CDC}{\breve{\cal D}}  \nc{\WD}{{\widehat{\cal D}}}
\nc{\nul}{{(0)}} \nc{\one}{{(1)}}
\nc{\aab}{\boldsymbol{\alpha}}  \nc{\abb}{\boldsymbol{\beta}}
\nc{\pj}{{\textstyle \prod_{j\in\CD}}\;\qq{j}}
\nc{\half}{{\textstyle\frac{1}{2}}}

\nc{\al}{\alpha}  \nc{\sig}{\sigma}   \nc{\ZN}{\mathbb{Z}_N}
\nc{\bg}{\boldsymbol{\gamma}} \nc{\bdr}{{\boldsymbol{\rho}}}
\nc{\bu}{{\bf u}} \nc{\bv}{{\bf v}}
\nc{\bV}{{\bf V}}   \nc{\bt}{{\bf t}}
\nc{\bnul}{{\mathbf 0}}
\nc{\bep}{\bu_{n+1}^{-1}(\ap-\bp\bv_{n+1})} \nc{\Xop}{\mathbf{X}}
\nc{\qq}[1]{e^{{\rm i}{\tilde q_{#1}}}}
\def\r#1{(\ref{#1})}
\nc{\qqq}[2]{e^{{\rm i}{\tilde q_{#1}}+{\rm i}{\tilde q_{#2}}}}
\nc{\sih}[2]{\sinh\frac{1}{2}(\g(#1)+\g(#2))}

\nc{\ra}{\rangle} \nc{\BAR}{\begin{array}} \nc{\EAR}{\end{array}}
\nc{\bdm}{\begin{displaymath}} \nc{\edm}{\end{displaymath}}
\nc{\be}{\begin{equation}} \nc{\ee}{\end{equation}}
\nc{\beq}{\begin{equation*}} \nc{\eeq}{\end{equation*}}
\nc{\ba}{\begin{array}} \nc{\ea}{\end{array}}
\nc{\bea}{\begin{eqnarray}} \nc{\eea}{\end{eqnarray}}

\nc{\hu}{{\bf u}}\nc{\hh}{{\hat
 h}}\nc{\bl}{\boldsymbol{\lambda}}\nc{\hv}{{\bf v}}
\nc\si{{\mathrm{s}}} \nc{\TS}{{\tilde S}} \nc{\A}{\mathcal{A}}
\nc{\Sc}{\mathcal{S}}  \nc{\N}{\mathcal{N}} \nc{\rL}{{\rm L}}
\nc{\rR}{{\rm R}} \nc{\fba}{{\zeta}} \nc{\qs}{Q} \nc{\lms}{s}
\nc{\av}{\prod_{s\,\in\,\ZN}}    \nc{\sga}{\sum_{\g\,\in\,\ZN}}
\def\sx#1{\sigma^x_{#1}} \def\sz#1{\sigma^z_{#1}}
\def\qu{{\sf q}}
\def\pu{{\sf p}}

\def\RRR{{\sf R}}
\nc\ve{{\varepsilon}}
\nc\mun{\nu}

\nc\sip{\gamma^\prime}\nc\ma{a'}\nc\mb{b'}\nc\mc{c\,'}\nc\md{d\,'}
\nc\pra{a''}\nc\prb{b''}\nc\prc{c\,''}\nc\prd{d\,''}
\nc{\txt}{\textstyle} \nc{\tq}{Q} \nc{\s}{{\gamma}}

\begin{document}
\title[Ising model spin matrix elements from Separation of Variables]
{Factorized finite-size Ising model spin matrix elements from
Separation of Variables}
\author{G von Gehlen$^\dag$,~ N Iorgov$^\ddag$,~ S~Pakuliak$^{\sharp\flat}$ and V~Shadura$^\ddag$}
\address{$^\dag$\ Physikalisches Institut der Universit\"at Bonn,
Nussallee 12, D-53115 Bonn, Germany}
\address{$^\ddag$ Bogolyubov Institute for Theoretical Physics, Kiev 03680,
Ukraine}
\address{$^\sharp$\ Bogoliubov Laboratory of Theoretical Physics,
Joint Institute for Nuclear Research, Dubna 141980, Moscow region,
Russia}
\address{$^\flat$\ Institute of Theoretical and Experimental Physics,
Moscow 117259, Russia}
\ead{gehlen@th.physik.uni-bonn.de,
iorgov@bitp.kiev.ua, pakuliak@theor.jinr.ru,
shadura@bitp.kiev.ua}
%

\begin{abstract}
Using the Sklyanin-Kharchev-Lebedev method of Separation of
Variables adapted to the cyclic Baxter--Bazhanov--Stroganov or
$\tau^{(2)}$-model, we derive factorized formulae for general
finite-size Ising model spin matrix elements, proving a recent
conjecture by Bugrij and Lisovyy.
\end{abstract}
\hspace*{2.5cm}{\small \today}\hspace*{12mm}   \submitto{\JPA}
\vspace*{-11mm} \pacs{75.10Hk, 75.10Jm, 05.50+q, 02.30Ik}

\section{Introduction}

Much work has been done on the 2-dimensional Ising model (IM)
during the past over 60 years. Many analytic results for the
partition function and correlations have been obtained. These have greatly contributed to
establish our present understanding of continuous phase
transitions in systems with short range interactions
\cite{Onsager,Yang,Kaufman,MPW,TwodimIsing,WMTB}.
Recent overviews with many references are given e.g. in \cite{Palm,BarryLec,Perk-AuYang}.
Many rather different mathematical approaches have been
used, so that already 30 years ago Baxter and Enting published the
``399th'' solution for the free energy \cite{BxE}, see also \cite{BaxAlg}. Spin-spin correlation
functions can be written as Pfaffians of Toeplitz determinants.
Most work has focussed on the thermodynamic limit and
scaling properties since these give contact to field theoretical
results and to beautiful Painlev{\'e} properties \cite{TwodimIsing,WMTB,BMcW}.

Only during the last decade more attention has been drawn to
correlations and spin matrix elements (form factors \cite{BKW}) in
finite-size Ising systems \cite{FZ,AKT,SE}.
Nanophysics experimental arrangements
often deal with systems where the finite size matters. Recent
theoretical work on the finite-size IM started from Pfaffians and
related Clifford approaches. In \cite{BL1} it has been pointed out that
one may write completely factorized closed expressions for
spin matrix elements of finite-size Ising systems. One goal of the
present paper is to prove the beautiful compact formula
conjectured in Eq.(12) of \cite{BL1}, see \r{ME_BL}. For achieving this, we introduce a method
which has not yet been applied to the Ising model: Separation of
Variables (SoV) for cyclic quantum spin systems. Our approach is
the adaption to cyclic models of the method introduced by Sklyanin \cite{Skly1,Skly2}
and further developed by Kharchev and Lebedev \cite{KarLeb1,KarLeb2}. We also make
extensive use of the analysis of quantum cyclic systems given in \cite{Tara}.

Little is known about state vectors of the 2-dimensional
finite-size IM. Only partial information about these state vectors
can be obtained from the work of \cite{Kaufman}. Recently Lisovyy \cite{Lisovyy}
found explicit expressions using the Grassmann algebra method. Here we shall present our
SoV approach \cite{gips,gipst1,gipst2} which gives explicit
formulas for finite-size state vectors
too. However, these come in a basis quite different from the one used in
\cite{Lisovyy}. We shall calculate spin matrix elements by directly
sandwiching the spin operator between state vectors. Factorized
expressions result if we manage to perform the multiple spin
summations over the intermediate states.

The prototype of a general $N$-state cyclic spin model is the
Baxter--Bazhanov--Stroganov model (BBS) \cite{BaxInv,BS,Kore},
also known as the $\tau^{(2)}$-model.
The standard IM is a very special degenerate
case of the BBS model. In order to avoid formulating many
precautions necessary when dealing with the very special IM, we
shall develop our version of the SoV machinery considering the
general BBS model. We chose to do this also because of the great
interest in the BBS model due to the fact that its transfer matrix
commutes with the integrable Chiral Potts model (CPM) \cite{BPAY,BaxtEV} transfer
matrix \cite{BS,BBP}. Obtaining state vectors for the CPM is a great
actual challenge \cite{AuYang-Eigenv,AuYang-Eigenv2}.
Although the eigenvectors for the transfer matrix of the BBS model
with periodic boundary condition are unknown for $N>2$, explicit formulas
for the eigenvectors of the BBS model with open and fixed boundary conditions have been found
\cite{Iorgov,IST}.

This paper is organized as follows:
In section 2 we define the BBS model and its Ising
specializations. In section 3 we discuss the Sklyanin SoV method adapted
to the BBS model as a cyclic system. We start with the necessary
first step, the solution of the associated auxiliary problem. In a second step we
obtain the eigenvectors and eigenvalues of the periodic system by
Baxter equations. The conditions which ensure that the Baxter equations have non-trivial solutions
are formulated as truncated functional equations. Section 4 gives a description
of local spin operators in terms of global elements of the monodromy matrix.
Starting with Section 5 we restrict ourselves to the case $N=2$, for which the BBS model
becomes a generalized 5-parameter plaquette Ising model. In section 6 we
further specialize to the homogenous case and then to the two-parameter Ising
case. Periodic boundary condition eigenvectors are explicitly constructed.
Section 7 is devoted to our main result, the proof of the factorized formula for Ising
spin matrix elements between arbitrary finite-size states. This is shown to agree with
the Bugrij-Lisovyy conjecture. In section 8 we give an analogous formula
for the Ising quantum chain in a transverse field. Finally, section 9 presents
our Conclusions. Large part of this paper relies on our
work in \cite{gips,gipst1,gipst2}.
Sections \ref{uvB}, {\ref{isin}} and \ref{fct} give new material.

\section{The BBS $\tau^{(2)}$-model}
\subsection{The inhomogenous BBS-model for general $N$}

We define the BBS-model as a quantum chain model. To each site $k$
of the quantum chain we associate a cyclic $L$-operator
\cite{BS,Kore} acting in a two-dimensional auxiliary space
\be\label{LBS}
L_k(\lm)=\left( \ba{ll} 1+\lm \kp_k \hv_k,\ & \lm
\hu_k^{-1} (a_k-b_k \hv_k)\\ [3mm] \hu_k (c_k-d_k \hv_k) ,& \lm
a_k c_k + \hv_k {b_k d_k}/{\kp_k } \ea \right)\!\!,\hspace*{4mm}
k=1,2,\ldots,n. \ee
$\lm$ is the spectral parameter, $n$ the
number of sites. There are five parameters $\kp_k$,
$a_k,\:b_k,\:c_k,\:d_k$ per site. $\hu_k$ and $\hv_k$ are elements
of an ultra local Weyl algebra, obeying
\beq
\fl \bu_j \bu_k=\bu_k \bu_j\,, \quad\;\bv_j \bv_k=\bv_k
   \bv_j\,,\quad\; \bu_j \bv_k=\om^{\delta_{j,k}}\bv_k\bu_j\,,\quad\;
   \om=e^{2\pi i/N}, \quad \bu_k^N=\bv_k^N=1. \eeq
At each site $k$
we define a $N$-dimensional linear space (quantum space) ${\cal
V}_k$ with the basis $ |\g\ra_k$, $\g\in \ZN$,  the dual space
${\cal V}_k^*$ with the basis $\: _k\langle\g|$, $\g\in \ZN$, and
the natural pairing $\: _k\langle\g'|\g\ra_k=\delta_{\g',\g}.$ In
$\;{\cal V}_k\:$ and $\;{\cal V}_k^*\:$ the Weyl elements $\bu_k$
and $\bv_k$ act by the formulas:
\be\label{uv} \fl \bu_k
|\g\ra_k=\om^\g |\g\ra_k\, ,\hx \bv_k |\g\ra_k = |\g+1\ra_k;\hq
_k\langle\g| \bu_k = \: _k\langle\g|\;\om^\g\ ,\hx \: _k\langle\g|
\bv_k  = \: _k\langle\g-1|\, . \ee
The monodromy  ${T}_n(\lm)$ and
transfer matrix ${\bf t}_{n}(\lambda)$ for the $n$ sites chain are
defined as \be\fl \label{mm} {T}_n(\lm)=L_1(\lm)\,\cdots L_n(\lm)=
\lk\!\! \ba{ll}
A_n(\lm)& B_n(\lm)\\[.5mm] C_n(\lm)& D_n(\lm) \ea \!\!\rk, \hx
{\bf t}_{n}(\lambda)=\mbox{tr}\: T_n(\lm)=A_n(\lm)+D_n(\lm).\,\ee
This quantum chain is integrable since the $L$-operators \r{LBS}
are intertwined by the twisted 6-vertex $R$-matrix at root of
unity \be
{R}(\la,\nu)\;=\;\lk\begin{array}{cccc} \la-\om\nu & 0 & 0 & 0 \\
[0.3mm] 0 & \om(\la-\nu) & \la (1-\om) & 0 \\ [0.3mm] 0 & \nu
(1-\om) & \la-\nu & 0 \\ [0.3mm] 0 & 0 & 0 & \la-\om\nu
\end{array}\rk\!, \ee
\be R(\la,\nu)\:  L^{(1)}_k\: (\la)
L^{(2)}_k(\nu)\;=\; L^{(2)}_k(\nu)\: L^{(1)}_k(\la)\: R(\la,\nu),
\label{rll} \ee where $\:L^{(1)}_k(\la)=
L_k(\la)\otimes\mathbb{I}$, $L^{(2)}_k(\la)=\mathbb{I} \otimes
L_k(\la)$. Relation (\ref{rll}) leads to  $\:[{\bf
t}_{n}(\lambda),{\bf t}_{n}(\mu)]=0\,.$ So ${\bf t}_n(\la)$ is the
generating function for the commuting set of non-local and
non-hermitian Hamiltonians $\;{\bf H}_0,\ldots,{\bf H}_n$:
\be\label{tr_mat_bs} {\bf t}_n(\lm) \:=\:{\bf H}_0\,+{\bf
H}_1\lm\,+\cdots\,+{\bf H}_{n-1}\lm^{n-1}\,+{\bf H}_{n}\lm^{n}.
\ee From \r{rll} it also follows that the upper-right entry
$B_n(\lm)$ of ${T}_n(\lm)$ is the generating function for another
commuting set of operators $\;{\bf h}_1,\ldots,{\bf h}_n$:
\be\label{Bpol} \left[ \,B_n(\lambda),\, B_n(\mu)\,\right]\:=\:0,\qquad
B_n(\lm)\,=\: {\bf h}_1 \lm \,+\, {\bf h}_2 \lm^2\, +\cdots\, +\, {\bf
h}_{n}\lm^{n}\,  . \ee Observe that ${\bf H}_0$ and ${\bf H}_n$
can be easily written
explicitly in terms of the global $\ZN$-charge rotation operator
$\;\bV_n$
\be \label{h0}\fl {\bf
H}_0\,=\,1\,+\bV_n\prod_{k=1}^n\frac{b_k d_k}{\kp_k}\,,\qquad {\bf
H}_n\,=\,\prod_{k=1}^n\: a_k\, c_k\,+\,\bV_n\prod_{k=1}^n
\kp_k\,,\qquad \bV_n\;=\;\bv_1 \bv_2 \cdots \bv_n.
\label{Znc} \ee
Here we shall not explain the great interest in the BBS-model due
to a second intertwining relation in the Weyl-space indices found in \cite{BS}
 and the related fact that for particular
parameters the Baxter $Q$-operator of the BBS-model is the
transfer matrix of the integrable Chiral Potts model, see
\cite{BS,BBP,RoanQ}. We will also not discuss the generalizations of the BBS model
introduced by Baxter in \cite{Btau},
and not explain how \r{LBS} arises in cyclic representations of the quantum group $U_q(sl_2)$, see
e.g. \cite{Tara,TarasovL,RoanUq}.

The transfer matrix \r{mm} can be written equivalently as a
product over face Boltzmann weights \cite{BaxInv,BBP}:\\[1mm]
${\bf t}_{n}(\lambda)\:=\:\prod_{k=2}^{n+1}\;
 W_\tau(\g'_{k-1},\g'_k,\g_k,\g_{k-1})\;\;$
with the face Boltzmann weights \bea\fl
 \lefteqn{W_\tau(\g_{k-1},\:\g_{k},\:\sip_{k-1},\:\sip_{k})=
{\textstyle\sum_{m_{k-1}=0}^1}\omega^{m_{k-1}(\sip_{k}
-\s_{k-1})}}\ny\\[2mm] \fl\hs\hs\times\:(-\omega t_q)^{ \s_{k} -
\sip_{k}- m_{k-1}}F'_{k-1}( \s_{k-1}- \sip_{k-1},m_{k-1})\;
   F''_{k}( \s_{k}- \sip_{k}, m_{k-1})\label{BW}\eea
where $m_k\in\lb 0,1\rb$ and $F_k'(\Delta\g,m_k)=F_k''(\Delta\g,m_k)=0$ if $\Delta\g\neq \lb 0,1\rb$,
and the non-vanishing values are \be
F_k'=\lk\ba{cc} 1 & \lm\: a_k\\ \kp_k & -b_k/\omega \ea\rk,
        \hx F_k''=\lk\ba{cc} 1 & \lm\: c_k\\ 1 & -d_k/\kp_k
        \ea\rk.\ee
The vanishing of $\:F_k'(\Delta\g,m_k)\:$ and $\:F_k''(\Delta\g,m_k)\:$ for $\:\Delta\g\neq \lb
0,1\rb\:$ means that the vertically neighboring $\ZN$-spins cannot
differ by more than one. The equivalence to the transfer matrix
defined by \r{LBS} and \r{mm} is seen writing the matrix elements of \r{LBS}
as\\[-4mm]
\be \fl \langle \g'_k|
L_k(\lm)_{m_{k-1},m_k}|\g_k\ra=\omega^{m_{k-1}\g'_k-m_{k}\g_k}\la^{\g'_{k}-\g_{k}-m_{k-1}}
F''_{k}( \g'_{k}- \g_{k}, m_{k-1}) F'_{k }( \g'_{k }- \g_{k },
m_k).\ee
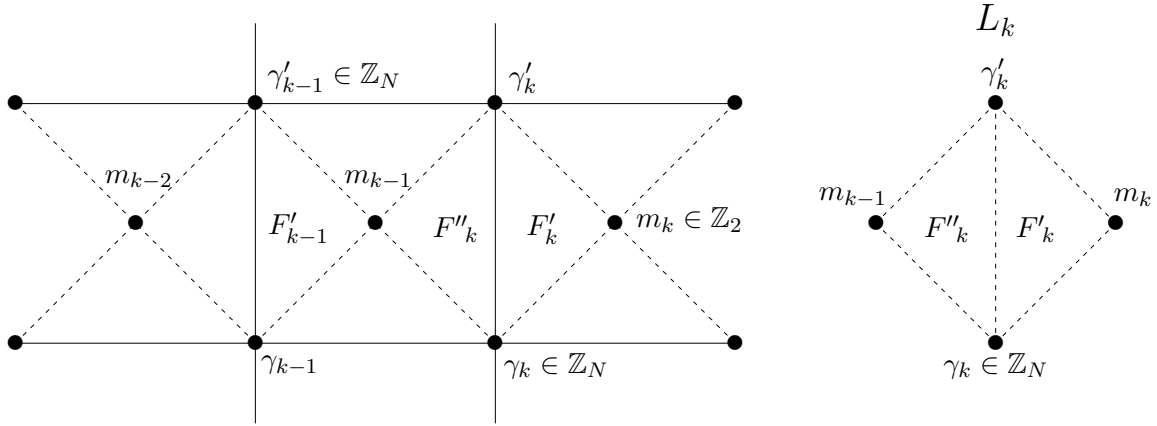
\begin{figure}[h]
\begin{center}
\renewcommand{\dashlinestretch}{30}
\unitlength=0.0336pt
\begin{picture}(8300,4300)(2300,450)
\drawline(2715,3612)(5415,3612)(5415,912)
    (2715,912)(2715,3612)
\drawline(15,3612)(2715,3612)(2715,4512)
\drawline(5415,4512)(5415,3612)(8115,3612)
\drawline(15,912)(2715,912)(2715,12)
\drawline(5415,12)(5415,912)(8115,912)
\dashline{60.000}(15,3612)(2715,912)(5415,3612)(8115,912)
\dashline{60.000}(15,912)(2715,3612)(5415,912)(8115,3612) \large
\put(4100,2750){\makebox(0,0)[cc]{{\small$m_{k-1}$}}}
\put(7600,2300){\makebox(0,0)[cc]{\small$m_k\in\mathbb{Z}_2$}}
\put(1400,2750){\makebox(0,0)[cc]{{\small$m_{k-2}$}}}
\put(3100,660){\makebox(0,0)[cc]{\small$\g_{k-1}$}}
\put(3200,2200){\makebox(0,0)[cc]{\small$F'_{k-1}$}}
\put(5950,2200){\makebox(0,0)[cc]{\small$F'_k$}}
\put(4970,2200){\makebox(0,0)[cc]{\small${F''}_{\!k}$}}
\put(6100,660){\makebox(0,0)[cc]{\small$\g_k\in\mathbb{Z}_N$}}
\put(3600,3900){\makebox(0,0)[cc]{\small$\g'_{k-1}\in\mathbb{Z}_N$}}
\put(5730,3900){\makebox(0,0)[cc]{\small$\g'_k$}}
\put(15,912){\makebox(0,0)[cc]{$\bullet$}}
\put(2715,3612){\makebox(0,0)[cc]{$\bullet$}}
\put(2715,912){\makebox(0,0)[cc]{$\bullet$}}
\put(5415,3612){\makebox(0,0)[cc]{$\bullet$}}
\put(5415,912){\makebox(0,0)[cc]{$\bullet$}}
\put(8115,3612){\makebox(0,0)[cc]{$\bullet$}}
\put(8115,912){\makebox(0,0)[cc]{$\bullet$}}
\put(15,3612){\makebox(0,0)[cc]{$\bullet$}}
\put(1365,2262){\makebox(0,0)[cc]{$\bullet$}}
\put(4065,2262){\makebox(0,0)[cc]{$\bullet$}}
\put(6765,2262){\makebox(0,0)[cc]{$\bullet$}}
\put(9700,2262){\makebox(0,0)[cc]{$\bullet$}}
\put(11050,912){\makebox(0,0)[cc]{$\bullet$}}
\put(11050,3612){\makebox(0,0)[cc]{$\bullet$}}
\put(12400,2262){\makebox(0,0)[cc]{$\bullet$}}
\dashline{60.0}(11050,912)(12400,2262)(11050,3612)(9700,2262)(11050,912)(11050,3612)
\put(11050,660){\makebox(0,0)[cc]{\small$\g_k\in\mathbb{Z}_N$}}
\put(11050,3950){\makebox(0,0)[cc]{\small$\g'_k$}}
\put(9440,2550){\makebox(0,0)[cc]{\small$m_{k-1}$}}
\put(12600,2550){\makebox(0,0)[cc]{\small$m_k$}}
\put(10500,2200){\makebox(0,0)[cc]{\small${F''}_{\!k}$}}
\put(11500,2200){\makebox(0,0)[cc]{\small${F'}_{\!k}$}}
\put(11050,4550){\makebox(0,0)[cc]{\large$L_k$}}
\end{picture}
\end{center}
\caption{Illustration of the two versions: left we see the $W$ of
\r{BW} indicated by full lines, whereas \r{mm} the $L_k$ of
\r{LBS} arise if we look at the lattice formed by the dashed lines
in the left figure and the dashed rhombus shown at the right. The
lattice is built by $\ZN$-spins on the full lines and
$\mathbb{Z}_2$-spins in the centers.}
\end{figure}

\subsection{Homogenous BBS-model for $N=2$}\label{isin}

The integrability of the BBS-model is valid also if the parameters
$\kp_k,\;a_k, \ldots,d_k\:$ vary from site to site and the construction of
eigenvalues and eigenvectors can be performed for this general
case. However, in order to obtain compact explicit formulas for
matrix elements, we shall often put all parameters equal:
$\kp_k\,=\,\kp,\ldots,\:d_k\,=\,d\,$ and call this the homogenous model.
In \cite{BIS} it has been shown that for
$N=2$ the general homogenous BBS-model can be rewritten as a generalized
plaquette Ising model with Boltzmann weights
\be
W(\sig_1,\sig_2,\sig_3,\sig_4)\;=\;a_0\lk 1\, +\,
{\textstyle\sum_{1\le i< j\le 4}}
          \,a_{ij}\,\sig_i\sig_j\,+\,a_4\,\sig_1\sig_2\sig_3\sig_4\rk,
          \label{FFIsing}\ee
subject to the free-fermion condition
$\;a_4=a_{12}a_{34}-a_{13}a_{24}+a_{14}a_{23}.$

For $N=2$ the Weyl elements can be represented by Pauli matrices.
Fixing $\kp=1$ the $L$-operator becomes
\[
L_k(\lm)=\lk\ba{cc} 1\,+\lm\,\sx{k} &  \lm \,\sz{k}\, (a\,-b\, \sx{k})\\[2mm]
\sz{k}\, (c\,-d\, \sx{k}) & \lm a c \,+ \sx{k}\, b\,d\ea \rk
\]
degenerating at $\lm\,=\,b/a$:
\[  L_k(b/a)\;=\;\lk\ba{c} 1\,+\, b/a\: \sx{k}\\ \sz{k} (c-d \sx{k}) \ea \rk\lk
\ba{cc}\!\! 1\,,&\!\! b\,\sz{k}\!\ea \rk\!. \]
The matrix elements of the corresponding transfer-matrix are
\[\fl
\langle \{\sigma'\}\,|\,{\bf t}_n(b/a)\,|\, \{\sigma\}\ra\;=\;
\prod_{k=1}^n \left(\delta_{\sigma_k,\sigma'_k} (1\,+\,b\,c\, \sigma_{k-1}\, \sigma'_k)
+\delta_{\sigma_k,-\sigma'_k}\; b/a\, (1\,-\,a\,d\, \sigma_{k-1}\, \sigma'_k)\right),
\]
where $\{\sigma\}=\{\sigma_1,\ldots,\sigma_n\}$ and $\{\sigma'\}=\{\sigma'_1,\ldots,\sigma'_n\}$
are the values of the spin variables of two neighboring rows,
$\sigma_k=(-1)^{\gamma_k}$, $\sigma'_k=(-1)^{\gamma'_k}\in \{+1,-1\}$,
and the identifications $\sigma_{n+k}\,=\,\sigma_k$, \;$\sigma'_{n+k}\,=\,\sigma'_k$ are used.

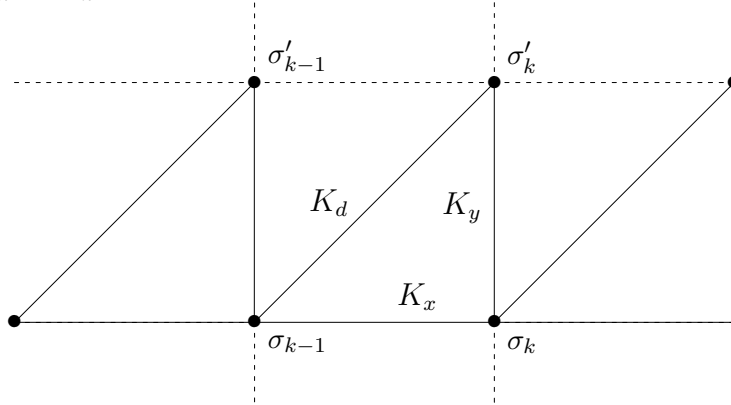
\begin{figure}[h]
\begin{center}
\renewcommand{\dashlinestretch}{30}
\unitlength=0.0336pt
\begin{picture}(5300,3600)(1300,450)
\dashline{60.000}(15,3612)(2715,3612)(2715,4512)
\dashline{60.000}(5415,4512)(5415,3612)(8115,3612)
\dashline{60.000}(15,912)(2715,912)(2715,12)
\dashline{60.000}(5415,12)(5415,912)(8115,912)
\drawline(15,912)(2715,3612)(2715,912)(5415,3612)(5415,912)(8115,3612)
\drawline(15,912)(8115,912)
\dashline{60.000}(2715,3612)(5415,3612)
\put(3200,660){\makebox(0,0)[cc]{\small$\sigma_{k-1}$}}
\put(5730,660){\makebox(0,0)[cc]{\small$\sigma_k$}}
\put(3200,3900){\makebox(0,0)[cc]{\small$\sigma'_{k-1}$}}
\put(5730,3900){\makebox(0,0)[cc]{\small$\sigma'_k$}}
\put(15,912){\makebox(0,0)[cc]{$\bullet$}}
\put(2715,3612){\makebox(0,0)[cc]{$\bullet$}}
\put(2715,912){\makebox(0,0)[cc]{$\bullet$}}
\put(5415,3612){\makebox(0,0)[cc]{$\bullet$}}
\put(5415,912){\makebox(0,0)[cc]{$\bullet$}}
\put(8115,3612){\makebox(0,0)[cc]{$\bullet$}}
\put(3550,2262){\makebox(0,0)[cc]{$K_d$}}
\put(4550,1200){\makebox(0,0)[cc]{$K_x$}}
\put(5050,2200){\makebox(0,0)[cc]{$K_y$}}
\end{picture}
\end{center}
\caption{Transfer-matrix for the triangular Ising lattice. Solid lines show the interaction
between spins.}\label{triang-fig}
\end{figure}

The matrix elements of the transfer-matrix of the Ising model on the triangular lattice
(see Fig.~\ref{triang-fig}) are
\be\label{t_triangle}
\fl \langle \{\sigma'\}|{\bf t}_\triangle| \{\sigma\}\ra\;=\;
\prod_{k=1}^n\:\:\exp(K_x\: \sigma_{k-1} \sigma_k+ K_y\: \sigma_{k} \sigma_k'+K_d\: \sigma_{k-1} \sigma'_k)\,.
\ee
The $k$-th factor of this product, taken at $\sigma_{k}= \sigma_k'$ is
\bea
\lefteqn{\exp(K_y)\:\exp((K_x+K_d)\: \sigma_{k-1}\: \sigma'_k)\:=}\ny\\
 &&\hs\hs=\;\exp(K_y)\:\cosh(K_x\,+\,K_d)\lk 1\,+\,\tanh(K_x\,+\,K_d)\: \sigma_{k-1}\: \sigma'_k\rk,
\ny\eea
and at $\sigma_{k}= -\sigma_k'$ is
\bea  \lefteqn{
\exp(-K_y)\:\exp((K_d-K_x)\: \sigma_{k-1}\: \sigma'_k)=}\ny\\
 &&\hs\hs=\:\exp(-K_y)\:\cosh(K_d-K_x)\lk 1\,+\,\tanh(K_d-K_x)\: \sigma_{k-1}\: \sigma'_k\rk\,.\ny
\eea
Now it is easy to compare the transfer-matrices $\:{\bf t}_n(b/a)\:$ and $\:{\bf t}_\triangle\,$:
\[\fl
{\bf t}_\triangle=\exp(n K_y)\:\cosh^n(K_x+K_d)\,{\bf t}_n(b/a)\,,\qquad \exp(-2K_y)\;\frac{\cosh(K_d-K_x)}{\cosh(K_d+K_x)}\:=\:b/a\,,
\]\[\fl
\tanh(K_x+K_d)\,=\,b\,c\,,\qquad \tanh(K_x-K_d)\,=\,a\,d\,.
\]
Although we considered ${\bf t}_n(\lm)$ at the special value of the spectral parameter $\lm=b/a$,
the transfer-matrix eigenstates are independent of this choice of $\lm$. So
the  eigenstates of the transfer-matrix of the Ising model on the triangular lattice
appear as eigenstates of the general homogenous BBS-model for $N=2$
(the parameter $\kp$ and one of the parameters $a$, ..., $d$ in the case of
homogeneous periodic BBS model can be absorbed by a rescaling of the other parameters
and using a diagonal similarity transformation
of the $L$-operators). The formulas for them will be
given later. Unfortunately, factorized formulas
for the matrix elements of the spin operator in this general
case have not been found. There are only two special cases for which such formulas are available:

\begin{itemize}
\item Row-to-row transfer-matrix for the Ising model on the square lattice:

\be\label{ISI}
\fl a=c,\quad b=d:\qquad\quad K_d=0,\qquad e^{-2K_y}\,=\,{b}/{a}\,,\quad\; \tanh K_x\,=\,a\,b\,.
\ee

This case will be the main object of our attention. It is the most general case where we have the
factorized formula for the spin operator matrix elements found by Bugrij and Lisovyy.

\item Diagonal-to-diagonal transfer-matrix for the Ising model on the square lattice:
\be\label{ISIdiag}
\fl a=c,\quad b=-d:\qquad\quad K_x=0,\quad e^{-2K_y}\,=\,{b}/{a}\,,\quad\; \tanh K_d\,=\,a\,b\,.
\ee

It is known \cite{Baxter:book}
that such transfer-matrices with different parameters $K_y\,=\,L$, $\,K_d\,=\,K$
(and corresponding $a$,$\,b$) constitute a commuting set of matrices having common eigenvectors provided
\be \label{modulekp}
\sinh 2K\;\, \sinh 2L\;=\;\frac{a^2\,-\,b^2}{1\,-a^2\,b^2}\;=\;\frac{1}{{\sf k}'}
\ee
is fixed. Thus in this case the eigenvectors depend
 on ${\sf k}'$ only. Therefore, in order to find the eigenvectors and the corresponding matrix elements
 of the spin operator it is
sufficient to fix $a\,=\,c\,=1/({\sf k}')^{1/2}$ and  $\:b\,=\,d\,=\,0$ and so to obtain a special case of the formulas
for the row-to-row transfer-matrix of the Ising model on the square lattice. Note that we get \cite{gipst2} the same
matrix elements in the case of the quantum Ising chain in a transverse field with strength $\:{\sf k}'\,$
because the corresponding Hamiltonian commutes with the transfer-matrices having the same $\:{\sf k}'$.
Another remark: with the restriction $a=c$, $b=-d$, $\kp=1$,
the transfer-matrices commute among themselves at independent values of {\it two} spectral parameters:
$\lm$ and the parameter which uniformizes \r{modulekp} (a parameter on the elliptic curve with modulus ${\sf k}'$,
see \cite{Baxter:book}).

\end{itemize}

\section{Separation of Variables for the cyclic BBS-model}
\subsection{Solving the auxiliary system \r{Bpol}: Eigenvalues and eigenvectors of $B_n(\lm)$.}

We start giving a summary of the SoV method as applied to the
general inhomogenous $\ZN$-BBS model \cite{gips}. The aim is to
find the eigenvalues and eigenstates of the $n$-site periodic
transfer matrix ${\bf t}_{n}(\lambda)$ of \r{mm}, and the idea
\cite{Skly1,Skly2,KarLeb1,KarLeb2} is first to construct a basis of the
$N^n$-dimensional eigenspace from eigenstates of $B_n(\lm)$, see
\r{Bpol}. This can be done by a recurrent procedure. Then the
eigenstates of ${\bf t}_{n}(\lambda)$ are written as linear
combinations of the $B_n(\lm)$-eigenstates. The multi-variable
coefficients are determined by Baxter $T-Q$-equations which by SoV
separate into a set of single-variable equations.

{}From \r{Bpol} the eigenvalues of $B_n(\lm)$ are polynomials in the
spectral variable $\lm$. Factorizing this polynomial, for $n\ge 2$  we get
\be\label{Bfact} B_n(\lambda)\,|\Psi_{\bl}\ra\:
=\:\lm\lm_0\prod_{k=1}^{n-1}\,(\lm-\lm_k)\,|\Psi_{\bl}\ra; \qquad
{\bl}=\{\lambda_0, \lambda_1,\ldots,\lambda_{n-1}\},\ee where
$\lm_1,\lm_2,\ldots,\lambda_{n-1}$ are the $n-1$ zeros of the
eigenvalue polynomial and $\lm_0$ is a normalizing factor. We can
label the eigenvectors by $\bl$.

An overview of the space of eigenstates of $B_n(\lm)$ is easily
obtained using the intertwining relations \r{rll}. It follows from
\r{rll} that the operators $A_n(\lm)$ and $D_n(\lm)$ of the
monodromy \r{mm}, taken at a zero $\lm=\lm_k$, are cyclic ladder
operators with respect to the $k$th component of $\bl$ in
$|\Psi_{\bl}\ra$. To see this consider e.g. the intertwining relation
\be\fl
(\lambda-\omega\mu)A_n(\lambda)B_n(\mu)\:=\:\omega(\lambda-\mu)
B_n(\mu)A_n(\lambda)+\mu(1-\omega)A_n(\mu)B_n(\lambda)\label{BA}
\ee which is a component of \r{rll}. Fixing $\;\lm=\lm_k$,
$\;k=1,\ldots, n-1\:$, in  (\ref{BA}) and acting on $\Psi_{\bl}$,
the last term in \r{BA} vanishes and we obtain \be
B_n(\mu)\left(A_n(\lambda_k)\:|\Psi_{\bl}\ra \right)\,=\mu\,\lambda_0\, (
\mu-{\omega}^{-1} {\lambda_k})\: {\textstyle \prod_{s\neq k}} (
\mu-\lambda_s)\: \left(A_n(\lm_k)\:|\Psi_{\bl}\ra\right)\!.\ee
This means that
\be\label{alm} A_n(\lambda_k)\,|\Psi_{\bl}\ra\:=\:\varphi_k\cdot
|\Psi_{\lambda_0\,, \,\ldots\, ,\: {\omega}^{-1} \lambda_k ,\, \ldots \,,\,
\lambda_{n-1}}\ra\,.\ee
Later we shall give an explicit expression
for the proportionality factor $\varphi_k$. Similarly, from another
component of \r{rll} and with another factor
    $\widetilde{\varphi}_k$ we get
\be\label{dlm}
D_n(\lambda_k)|\Psi_{\bl}\ra\:=\:\widetilde{\varphi}_k\cdot
|\Psi_{{\omega}^{-1}\lambda_0\, ,\, \ldots\, ,
 \:\omega\, \lambda_k\, ,\, \ldots , \,\lambda_{n-1} }\ra.
\ee Furthermore, acting by \r{BA} on $|\Psi_{\bl}\ra$ and extracting
the coefficient of $\;\lm^{n+1}\mu^n\;$ we get
\be\label{vvv} \bV_n
|\Psi_{\bl}\ra\:=\:|\Psi_{ {\omega}^{-1}\lambda_0\, ,
 \: \lambda_1\, ,\, \ldots\, , \,\lambda_{n-1} }\ra\,.\ee
Assuming generic parameters in $L_k$ such that all
proportionality factors are non-vanishing, by repeated application
of $A_n(\lm_k),\: D_n(\lm_k)\:\mbox{and}\: \bV_n$ to any
eigenstate $|\Psi_{\bl}\ra$ we span the whole $N^n$-dimensional space
of states. Later, when we give explicit expressions for
$\varphi_k$ and $\widetilde{\varphi}_k$ we can check whether these factors
can vanish.

 So, if for a given set of parameters
$a_k,b_k,c_k,d_k,\kp_k,\:(k=1,\ldots,n)$ there is an
eigenvector with the eigenvalue polynomial determined by the zeros
$\bl$, then there are also eigenvectors to all eigenvalue
polynomials determined by the zeros
\be\{\lm_0\,\om^{\rho_{n,1}},\,\ldots\,,\lm_{n-1}\,\om^{\rho_{n,n-1}}\}
\;\;\:\mbox{with}\;\;\:
\bdr_n=(\rho_{n,0},\ldots,\rho_{n,n-1})\in(\ZN)^n.\label{br} \ee
Let us therefore write the zeros as
\be \lm_{n,k}\;=\;-r_{n,k}\,\om^{\rho_{n,k}}, \label{lr}\ee
where for $n$ fixed,
the $n$ real numbers $r_{n,k}$ are determined by the $5n$ parameters
$\,a_l,\ldots,\kp_l$. For fixed parameters the $N^n$ in all following calculations
we shall label the eigenvectors by the $\bdr_n$ instead of our
previous $\lm_{n,k}$. For given parameters, the set of the eigenvalues
is determined by the $r_{n,0},\ldots,r_{n,n-1}$. The eigenvalue equation
for $B_n(\lm)$ becomes
\be\label{Be}
B_n(\lm)|\Psi_{\bdr_n}\ra =\lm\, r_{n,0}\,\om^{-\rho_{n,0}}
\prod_{k=1}^{n-1}\lk\lm+r_{n,k}\om^{-\rho_{n,k}}\rk|\Psi_{\bdr_n}\ra\,,
\ee

In order to calculate the $r_{n,k}$ in terms of the parameters, we don't
 need the full quantum transfer matrix and the $L_k$-operators involving the Weyl variables.
 Rather, by the following averaging procedure \cite{Tara}
\be {\cal O}(\lm^N)\;=\;\langle\,O(\lm^N)\,\rangle=\;{\textstyle
\av} O(\om^s\lm).\label{ave}\ee
we associate to a spectral parameter dependent quantum operator $O(\lm)$
a classical counterpart $\,{\cal O}(\lm^N)\,.$
We define the classical BSS model by the $L$-operator $\mathcal{L}_m(\lm^N)$
\be \label{avL}
\fl \mathcal{L}_m(\lm^N)\;=\;\left(\ba{cc} \langle\, L_{00}\,\rangle&
         \langle\,L_{01}\,\rangle \\[2mm]
         \langle\,L_{10}\,\rangle & \langle\,L_{11}\,\rangle\ea\right)
=\left( \ba{cc}1-\epsilon \kp_m^N \lm^N &\;\; -\epsilon\lm^N(a_m^N-b_m^N)\\[2mm]
            c_m^N-d_m^N &\;\; b_m^N d_m^N/\kp_m^N-\epsilon \lm^N a_m^N c_m^N\ea \right)\ee
where $\epsilon=(-1)^N$. Analogously, we define the classical monodromy ${\cal T}_n$ by
\be \mathcal{T}_n\:=\;{\cal L}_1(\lm^N)\:
{\cal L}_2(\lm^N) \cdots\: {\cal L}_n(\lm^N)\:=\:\left(\ba{cc}
{\cal A}_n(\lm^N) &  {\cal B}_n(\lm^N)\\[1mm]
{\cal C}_n(\lm^N) &  {\cal D}_n(\lm^N) \ea\right)\label{ABCDcl}\ee
Proposition 1.5 of \cite{Tara} tells us that the
classical polynomials ${\cal A}_n(\lm^N)$, ${\cal B}_n(\lm^N)$,
${\cal C}_n(\lm^N)$ and ${\cal D}_n(\lm^N)$ are the averages of their counterparts in \r{mm}:
$\;\;{\cal A}_n(\lm^N)\:=\:\langle \:A_n(\lm)\:\rangle,\;$ etc.
So for $n\ge 2$ we have
\be\label{zerB}
{\cal B}_m(\lm^N)=(-\epsilon)^m \lm^N r_{m,0}^N \prod_{s=1}^{m-1}(\lm^N-
\epsilon\, r_{m,s}^N).\ee
It is easy to derive \cite{gips} a three-term recursion which expresses
${\cal B}_m(\lm^N)$ in terms of ${\cal B}_{m-1}(\lm^N)$ and  ${\cal B}_{m-2}(\lm^N)$.
Using the initial values ${\cal B}_1(\lm^N)=-\epsilon\, \lm^N r_1^N; \hx {\cal B}_0(\lm^N)=0$ and defining
$r_1^N=a_1^N-b_1^N$, this gives a $(n-1)th$-degree algebraic relation for the $r_{m,s}^N$.

For the homogenous model (the constants are taken to be site-independent) this can be replaced by just
a quadratic equation, see the Appendix of \cite{gips}.

\subsection{Solving the auxiliary system:
               Explicit construction of the eigenvectors of $B_n(\lm)$.}

The stepwise construction of the eigenvectors, starting with one-site, then two-site
as linear combination of products of
two one-site eigenvectors etc. is tedious because we have to go to 4 sites before
the general rule emerges.

Let us start finding the one-site right eigenvectors $|\psi_\rho\ra_1$ of $B_1(\lm)$
as linear combination of spin states $|\g\ra_1$, $\;\g\in \ZN$, writing
\be\label{psi1}  |\psi_{\rho}\ra_1=\sga w_{p}(\g\:-\:\rho)\:|\g\ra_1,\hq \rho\in \ZN\,.\ee
Applying on the left $B_1$ from \r{LBS} and on the right \r{Be}, we demand
\be\fl    \lm\:\hu_1^{-1} (a_1-b_1 \hv_1)\sga w_{p}(\g\:-\:\rho)\:|\g\ra_1\;=
    \;\lm\:r_{1,0}\,\om^{-\rho_{1,0}}\sga w_{p}(\g\:-\:\rho)\:|\g\ra_1.  \ee
Applying \r{uv} and shifting the left hand summation for the term with $\:|\g+1\ra_1$, we get
\be (a_1\:-\,r_{1,0}\om^{\g-\rho})\:w_{p}(\g\:-\:\rho)\;=\;b_1\:w_{p}(\g\:-\:\rho\:-1).\ee
This is a difference equation for the function $w_p(\g)$ \cite{BB}:
\be\fl \frac{w_p(\g)}{w_p(\g-1)}\:=\:\frac{y}{1\,-\,\om^\g\,x}\,;
  \qquad\quad w_p(0)=1\,;\qquad \g\in\ZN\,,\label{Fermat} \ee
where we have put $\;y\,=\,b_1/a_1,\;r_{1,0}\,=\,x\,a_1$ and chose the
   initial value $w_p(0)=1$. The cyclic property $\:w_p(\g)\:=\:w_p(\g\,+\,N)\:$
imposes the Fermat condition $\:x^N\,+\,y^N\,=\,1$ on the two-component vector $p\:=\:(x,\:y\,)$.
We indicate $p\:$ as a subscript on the functions $w_p(\g)$.
We shall consider the case of ``generic parameters'', so in particular we exclude the case $a_k^N\,-\, b_k^N\,=\,0$,
and the ``superintegrable'' case \be
a_k\;=\;\om^{-1}\,b_k\;=\;c_k\;=\;d_k\;=\;\kp_k\;=\:1, \label{super}
\ee since in the latter cases degenerations appear.

We write the analogous left eigenvector as
\be  {}_1\langle \psi_{\rho}|=
 \sga \frac{1}{w_p(\g-\rho-1)}\;\;_1\langle \g| ,\qquad \rho\in \ZN \ee
 with the same functions $w_p(\g)$, ~just now $\;p\,=\,(r_{1,0}/a_1,\;\om^{-1}b_1/a_1).$
The Fermat vector dependent functions $w_p(\g)$ play an important role for cyclic models.
They are root-of-unity analogues of the $q$-gamma function.

By a similar calculation, the two-site eigenvectors are found to be:
\be\fl \label{Psi2}
|\Psi_{\,\rho_{2,0},\:\rho_{2,1}\,}\ra=\!\!\!
\sum_{\rho_1,\,\rho_2\in \ZN} \: \!\frac{\om^{-(\rho_{2,0}+\rho_{2,1}-\rho_1)
 (\rho_{2,0} -\rho_2)}}
{w_{p_{\,2,\,0}}(\rho_{2,0}-\rho_1-1) w_{\tilde p_{\,2}}(\rho_{2,0}+\rho_{2,1}-\rho_2-1 )}\:\;
|\psi_{\rho_1}\ra_1\:\otimes\; |\psi_{\rho_2}\ra_2\,.
\ee
where $p_{2,\,0}=(x_{2,\,0},y_{2,\,0})$,
$ \tilde p_{2}=( \tilde x_{2},\tilde y_{2})$ and
 \be
 \hspace*{-1.5cm} x_{2,\,0}=a_2\,c_2\frac {r_1}{r_{2,\,0}}\,, \hq
 y_{2,\,0}=\kp_1\,\frac {r_2}{r_{2,\,0}}\,,\hq
 \tilde x_{2}= \frac {r_2}{r_{2,\,0}\, r_{2,\,1}}\,, \hq
\tilde y_{2}= \frac{b_2\,d_2}{\kp_2}\frac {r_1}{r_{2,\,0}\, r_{2,\,1}}\,.\label{ptwo}
 \ee
The condition that $p_{2,\,0}$ and $\tilde p_2$  are Fermat vectors determines $\;r_{2,0}$ and $r_{2,1}.$

The explicit formula for both the left- and right eigenvectors of $B_n(\lm)$ for
general number of sites $n$ has been proved by lengthy induction and is given in \cite{gips}.
A by-product of these calculations are the formulas for $\:\varphi_k\:$
and $\:\widetilde{\varphi}_k\:$ introduced in \r{alm},\r{dlm}:
\be  \fl \label{phik} A_n(\lm_{n,k})\,|\Psi_{\bdr_n}\ra=\varphi_k(\bdr_n')\,|\Psi_{\bdr_n^{+k}}\ra,\quad
\varphi_k(\bdr_n')\,=\,-\frac{\tilde r_{n-1}}{r_n}\;\om^{-\tilde
\rho_{n}+\rho_{n,0}}\,F_n(\lm_{n,k}/\om)\,\prod_{s=1}^{n-2}\;y_{n-1,s}^{n,k},
 \ee
 \be\fl
D_n(\lm_{n,k})|\Psi_{\bdr_n}\ra=\tilde\varphi_k(\bdr_n')\,|\Psi_{\bdr_n^{+0,-k}}\ra,\quad
 \tilde\varphi_k(\bdr'_n)=
-\frac{r_n}{\tilde r_{n-1}}\frac{\om^{\trh_n-\rho_{n,0}-1}}
{\prod_{s=1}^{n-2}y_{n-1,s}^{n,k}}\;\prod_{m=1}^{n-1}\;F_m(\lm_{n,k}), \label{Dlmk}
\ee
 \be F_n(\lm)\:=\:\lk\, b_n\,+\om a_n \,\kp_n \lm\rk\,
\lk \,\la\, c_n\,+d_n/\kp_n\, \rk. \label{qdet}\ee
On the left of \r{phik} and \r{Dlmk} the eigenvectors $\Psi_{\bdr_n}$ of $B_n(\la)$
are labeled by the vector
\be\bdr_n=(\rho_{n,0},\ldots,\rho_{n,n-1})\in(\ZN)^n.\ee
$\bdr_n^{\pm k}\;$ denotes the vector $\;\bdr_n\;$ in which $\rho_{n,k}$ is
replaced by $\rho_{n,k}\pm 1$:
\be\bdr_n^{\pm k}=(\rho_{n,0},\ldots,\rho_{n,k}\pm 1,\ldots,\rho_{n,n-1})$,~
$\;\;k=0,1,\ldots,n-1,\ee
\be \rt_n\,=\,r_{n,0}\,r_{n,1}\,\ldots\, r_{n,n-1}\hs\mbox{and}\hs\trh_n={\textstyle\sum_{k=0}^{n-1}}\rho_{n,k}.\ee
$\bdr_n'$ denotes the vector $\bdr_n$ without the component $\rho_{n,0}$:
\be \bdr_n'\:=\:(\rho_{n,1},\ldots,\rho_{n,n-1})\in(\ZN)^{n-1}. \label{rhoprime}\ee
The $y_{n-1,s}^{n,k}$ are components of a Fermat vector
   $\;p^{n,k}_{n-1,s}\,=\,(x^{n,k}_{n-1,s},y^{n,k}_{n-1,s})$ defined by
$x^{n,k}_{n-1,s}\,=\:r_{n,k}/r_{n-1,s}$,  see Section 2.4 of \cite{gips}.
The $\:F_m(\lm)\:$ which appears in \r{phik} and \r{Dlmk} is a factor of the quantum determinant:
\be   A_n(\om \lm)D_n(\lm)-C_n(\om \lm) B_n(\lm)
       \:=\:\bV_n\cdot\prod_{m=1}^n\;F_m(\lm),\label{qdetall}   \ee
{}From \r{LBS} we can read off directly the $\lm^0$- and $\lm^n$-coefficients of the polynomial $A_n(\lm)$:
\be  A_n(\lm)\;=\;1\;+\ldots\;+\kp_1\kp_2\ldots\kp_n\:\bV\:\lm^n\,.  \ee
Then using \r{phik},
the general action of $A_n(\lm)$ on $B_n$ eigenvectors can be written as an interpolation polynomial
\bea\fl
\lefteqn{A_n(\lm)|\Psi_{\bdr_n}\ra\:=\:\prod_{s=1}^{n-1} \lk
1-\frac{\lm}{\lm_{n,s}}\rk\:|\Psi_{\bdr_n}\ra +\lm \kp_1\cdots\kp_n\:
\prod_{s=1}^{n-1} (\lm-\lm_{n,s})\,|\Psi_{\bdr_n^{+0}}\ra\;+}\ny\\ &&
\hs\hs\hs+\;\sum_{k=1}^{n-1}\; \lk\prod_{s\ne k}
\frac{\lm-\lm_{n,s}}{\lm_{n,k}-\lm_{n,s}}\!\rk\!
\frac{\lm}{\lm_{n,k}}\;\varphi_k(\bdr'_n)\;|\Psi_{\bdr_n^{+k}}\ra\,. \label{Alm}\eea

Considerable effort is needed to present the norm of an arbitrary state vector $|\Psi_{\bdr_n}\ra$
in factorized form, since multiple sums over the intermediate indices have to be performed.
The norms are independent of the phase $\rho_{n,0}$ and their dependence
on $\bdr'_n$ is:
\be  \fl\langle \Psi_{\bdr_n}|\Psi_{\bdr_n}\rangle \;=\;
\frac{C_n}{\prod_{l<m}(\la_{n,l}-\la_{n,m})}\;=\;
\frac{C_n}{\prod_{l<m}(r_{n,m}\om^{-\rho_{n,m}}-r_{n,l}\om^{-\rho_{n,l}})}\,.
\label{norm}\ee
The normalizing factor $C_n$ is independent of $\bdr_n$ and can be written recursively \cite{gipst1}.
The two lowest values are:
\be C_1\;=\;\frac{N}{\om}\lk\frac{x_1}{y_1}\rk^{N-1},\hs C_2\;=\;C_1\frac{N^3}{\om}
\lk\frac{x_2}{y_2\:{\tilde y}_2\, y_{2,0}}\rk^{N-1}. \ee

\subsection{Periodic model: Baxter equation and truncated functional equations}
\label{periodic}

In the auxiliary problem we looked for eigenfunctions of $B_n$. $B_n$ does
not commute with $\bV_n$ \r{Znc}, see \r{vvv}: $\bV_n|\Psi_{\bdr_n}\rangle=|\Psi_{\bdr_n^{+0}}\rangle$.
Now we are looking for eigenfunctions of $\bt_n$
which commutes with $\bV_n$. By Fourier transformation in $\rho_{n,0}$ we build a basis diagonal in $\bV$,
where the Fourier transformed variable $\rho\in\mathbb{Z}_N$ is the total $\ZN$-charge.:
\be |\tilde\Psi_{\rho,\bdr'_n}\rangle={\textstyle \sum_{\rho_{n,0}\in\ZN}}\om^{-\rho\cdot\rho_{n,0}}
     |\Psi_{\bdr_n}\rangle,\qquad
\bV_n\: |\tilde\Psi_{\rho,\bdr'_n}\rangle \;=\;\om^\rho\: |\tilde\Psi_{\rho,\bdr'_n}\rangle\,.\label{tPsi}
\ee
We now write the eigenfunctions $|\Phi_{\rho,{\bf E}}\rangle$ of $\bt_n(\lm)$ as linear combination of
the $|\tilde\Psi_{\rho,\bdr'_n}\rangle$.
The eigenvalues of $\bt_n(\lm)$ on these states are again order $n$ polynomials in $\lm$:
\be \bt_n(\lm)|\Phi_{\rho,{\bf E}}\rangle\:\;=\;
  (E_0+E_1\lm+\cdots+E_{n-1}\lm^{n-1}+E_{n}\lm^{n}) |\Phi_{\rho,{\bf E}}\rangle  \label{tE}\ee
Since the values of $E_0$ and $E_n$ can be read off immediately from \r{Znc}:
\be
 E_0\,=1+\om^\rho {\textstyle\prod_{m=1}^n}\:b_m d_m/\kp_m,\quad
E_n\,=\,{\textstyle\prod_{m=1}^n}\: a_m c_m\,+\om^{\rho}\,{\textstyle\prod_{m=1}^n}\,\kp_m\,,
\label{Enn}\ee
we combine the remaining coefficients into a vector $\;{\bf E}\;=\;\{E_1,\ldots,E_{n-1}\}\,,$
and label the eigenvectors just by the charge $\rho$ and ${\bf E}$:
\be  \fl \bt_n(\lm)\;\,|\Phi_{\rho,{\bf E}}\rangle\:\;=\;
  t_n(\lm|\rho,{\bf E})\;\,|\Phi_{\rho,{\bf E}}\rangle,\hs   |\Phi_{\rho,{\bf E}}\rangle \:=\:
\sum_{\bdr'_n} \; {\cal Q}^{\rm R}(\bdr'_{n}|\,\rho,{\bf E})\;\:|\tilde\Psi_{\rho,\bdr'_n}\rangle\,. \label{EV}
\ee
Now, in order to achieve SoV of the multi-variable functions ${\cal Q}^{\rm R}$, we split off from
$\:{\cal Q}^{\rm R}(\bdr'_{n}|\,\rho,{\bf E})\:$ Sklyanin's separating factor:
\be \label{QR}
{\cal Q}^{\rm R} (\bdr'_{n}|\,\rho,{\bf E})=\frac{\prod_{k=1}^{n-1}\:
\tq_k^{\rm R}(\rho_{n,k})}{ \prod_{s,s'=1\atop (s\ne s')}^{n-1}
w_{p_{n,s}^{n,s'}}(\rho_{n,s}-\rho_{n,s'})}\,.
\ee
We shall not give the detailed calculation and just indicate the main mechanism.
We express $\bt_n(\lm)$ as
an interpolation polynomial through the zeros $\lm_{n,k}$ of $B_n(\lm)$:
\bea \fl \lefteqn{ (A_n(\lm)\,+\,D_n(\lm))|\tilde\Psi_{\rho,{\bdr'}_{n}}\ra
\;=\;\lb E_0\:\prod_{s=1}^{n-1} \lk 1\,-\frac{\la}{\la_{n,s}}\rk\,+\la\; E_n
  \prod_{s=1}^{n-1}(\lm-\lm_{n,s}) \rb\tilde\Psi_{\rho,\bdr'_{n}}\;+\hs\hs}\ny\\
\hspace*{-6mm}&+&\:\sum_{k=1}^{n-1}\;\lk\prod_{s\ne k} \frac{\lm-\lm_{n,s}}{\lm_{n,k}-\lm_{n,s}}\rk
\!\frac{\lm}{\lm_{n,k}}\lk\varphi_k(\bdr'_{n})\,|\tilde\Psi_{\rho,{\bdr'}_{n}^{+k}}\ra\!
           +\om^\rho\, \tilde\varphi_k(\bdr'_{n})\,|\tilde\Psi_{\rho,{\bdr'}_{n}^{-k}}\ra\rk.\label{ApD}\eea
When we evaluate \r{ApD} successively at the $n-1$ values $\lm=\lm_{n,k},\;k=1,\ldots,n-1$,
the terms on the right of the first line
of \r{ApD} do not contribute. Due to the Sklyanin-factor the brackets involving
the differences $\lm_{n,k}-\lm_{n,s}$ are made to cancel, leading to SoV. This results in the
$n-1$ single-variable  $\lm_{n,k}$ Baxter equations ($k=1,\ldots,n-1)$
\be\fl
t_n(\lm_{n,k}|\rho,{\bf E})\;\;\tq^{\rm R}_k(\rho_{n,k})=
\Delta_k^+(\lm_{n,k})\;\tq^{\rm R}_k(\rho_{n,k}+1)+\Delta_k^-(\om\lm_{n,k})\;
\tq^{\rm R}_k(\rho_{n,k}-1)\,.\label{BAXR}\ee
Starting from the left eigenvectors the analogous left Baxter equations are
\be\fl  t_n(\lm_{n,k}|\rho,{\bf E})\;\;\tq^{\rm L}_k(\rho_{n,k})=
\om^{n-1}\Delta_k^-(\lm_{n,k})\;\tq^{\rm L}_k(\rho_{n,k}+1)+\om^{1-n}\Delta_k^+(\om\lm_{n,k})\;
\tq^{\rm L}_k(\rho_{n,k}-1)\,,\label{BAXL}\ee
where we abbreviated
\be \fl\Delta_k^+(\lm)\:=\:(\om^\rho/\chi_k)\,(\lm/\om)^{1-n}\:
         \prod_{m=1}^{n-1}\,F_m(\lm/\om)\,,\qquad
\Delta_k^-(\lm)\:=\:\chi_k\:(\lm/\om)^{n-1}\:F_n(\lm/\om)\,.\label{Delta} \ee
$\chi_k$ collects several factors (partly arising from $\varphi_k$ and $\widetilde{\varphi}_k$)
determined by constants $\kp_k,a_k,\ldots,d_k$ alone.
Now note that the left hand side of \r{BAXR} more explicitly reads
\beq  \lk E_0 + {\textstyle\sum_{s=1}^{n-1}}E_s\,\lm_{n,k}^s\;+\;E_n\lm_{n,k}^n\rk\;
                \tq^{\rm R}_k(\rho_{n,k})\;=\;\ldots\eeq
where the ${\bf E}$ are unknown and have to be determined from the
   system of homogenous equations \r{BAXR}
together with the $n-1$ functions $\;Q_k^{\rm R}(\rho_{n,k})$. In order to have a non-trivial solution,
the coefficient determinants have to be degenerate. Fix a $k$, then from the determinant
we may get one relation among $E_0,\ldots,E_n$. All $n-1$ systems for different $k\;$ should
be sufficient to determine all components of ${\bf E}$.~ Fortunately, the condition
for non-trivial solutions to \r{BAXR} can be
written as well-known truncated functional equations:

Define $\:\tau^{(2)}(\lm)\:=\:t(\lm)$\footnote{This definition in \cite{BS} is the origin of calling the BBS model
the $\:\tau^{(2)}$-model} and construct a fusion hierarchy \cite{KS82,BBP} by setting $\tau^{(0)}(\lm)=0$,
$\tau^{(1)}(\lm)=1$, and
\be\fl \tau^{(j+1)}(\lm)\;=\;\tau^{(2)}(\om^{j-1}\lm)\,\tau^{(j)}(\lm)\:-\:
\om^{\rho}\,z(\om^{j-1}\lm)\, \tau^{(j-1)}(\lm),\qquad j=2,3,\ldots,
N, \label{rectau} \ee
where \\ [-11mm]
\be\label{zDelta}
 z(\lm)\;=\;\om^{-\rho}\: \Delta^+(\lm)\:\Delta^-(\lm)\;=\;{\textstyle\prod_{m=1}^{n}}\, F_m(\lm/\om).
\ee
Then it can be shown \cite{gips} that if $\;\tau^{(N+1)}(\lm)\;$ satisfies the truncation identity
\be\label{reltau}
\tau^{(N+1)}(\lm)\:-\:\om^{\rho}\,z(\lm)\, \tau^{(N-1)}(\om\lm)\;=\;
{\cal A}_n(\lm^N)\,+\,{\cal D}_n(\lm^N)\ee
with ${\cal A}_n(\lm^N)\,+\,{\cal D}_n(\lm^N)$ given in \r{ABCDcl}, then the system
\r{BAXR} has a non-trivial solution for all $k$. This truncated hierarchy can be used to find
the transfer matrix eigenvalues \cite{KR87,BR89}.
In our construction we have even more: for every solution of \r{rectau},\r{reltau}
we can construct an eigenvector.

\section{Action of $\:\bu_k \,$ and $\:\bv_k\,$ on eigenstates of $\:B_n(\lm)$}\label{uvB}

Our main aim is to calculate matrix elements of the local operators $\bu_k$ and $\bv_k$ between
eigenstates $|\Phi_{\rho,{\bf E}}\rangle$ of $\:\bt_n(\lm)\,$. Since we know how to get these states from
the $B_n(\lm)$ eigenstates \r{EV},\r{QR},\r{BAXR}, we first set out to find the action of the
local operators on the $ |\Psi_{\bdr_n}\rangle$. Since we built our auxiliary states successively from
one-site to $n$-site, the formulas will not be symmetric between e.g. $\bu_j$ and $\bu_k$ with $j\ne k$.

For $\bu_n$ we can calculate its action directly. Starting from
\[
\bu_n^{-1} (a_n-b_n \bv_n)|  \psi_{\rho_n}\rangle_n\;=\;
r_n \om^{-\rho_n}|  \psi_{\rho_n}\rangle_n\,,
\]
 we get the formula for the action of $\bu_n$ on one-site eigenvectors:
\be\label{unpsi}
\bu_n |  \psi_{\rho_n}\rangle_n\;=\;\frac{\om^{\rho_n}}{r_n}\left(a_n |  \psi_{\rho_n}\rangle_n
- b_n |  \psi_{\rho_n+1}\rangle_n\right).
\ee
Using then the explicit recursion formula relating $\;\, |\Psi_{\bdr_n}\rangle\:$ to $\: |\Psi_{\bdr_{n-1}}\rangle\:$
one finds \cite{gipst1}:
\bea
\lefteqn{\bu_n\: | \Psi_{\bdr_n}\rangle=
 \frac{a_n}{\tilde r_n \om^{-\tilde \rho_n}}\; | \Psi_{\bdr_n}\rangle-
\frac{b_n \kp_1 \kp_2 \cdots \kp_{n-1}}{r_{n,0}\om^{-\rho_{n,0}}}\:
   | \Psi_{\bdr_n^{+0}}\rangle\;+}\label{unPsi}\\
&&\hs\;+\;\:\sum_{k=1}^{n-1}\;
\frac{a_n\, b_n\:\varphi_k({\bdr\,}'_n)}{r_{n,0}\om^{-\rho_{n,0}}\:\lm_{n,k}\:
  (b_n+a_n\kp_n\lm_{n,k})\:\prod_{s\ne k} (\lm_{n,k}-\lm_{n,s})}\;
  |\Psi_{\bdr_n^{+k}}\rangle\,.\ny \eea
We shall derive this result in a simpler way expressing the local operators $\bu_k$ and $\bv_k$ in terms
of the global entries $A_n$ and $B_n$ of monodromy matrix, taken at particular values of $\lm$.
There is  a well-known method elaborated by the Lyon group \cite{MT}. However, this method
requires the fulfillment of the condition $R(0)\:=\:P$ with $R$ the quantum $R$-matrix
intertwining two $L$-operators in quantum spaces
and $P$ the permutation operator. This requirement is fulfilled for the cyclic $L$-operators only
at special values of parameters where the $R$-matrix is the product of four weights of the Chiral Potts
model \cite{BS}. Another requirement regards the possibility to obtain such a $R$-matrix
by fusion in the auxiliary space of the initial $L$-operator. This requirement can not be fulfilled
for the cyclic $L$-operators \r{LBS} because the fusion in the auxiliary space \cite{Roan-fus} gives
$L$-operators with the highest weight evaluation representations of the corresponding quantum affine algebra,
but we need cyclic type representation in the auxiliary space.

We will use an idea borrowed from a paper of Kuznetsov on SoV for classical systems \cite{Kuz}.
What we can do is the following:
Consider the inverse of the operator $L_k(\lm)$:
\be  \fl L_k^{-1}(\lm)=\left( \ba{ll}
\om\,\lm\,a_k\,c_k\,+\,\hv_k\,{b_k\,d_k}/{\kp_k}\;\;&-\lm\,\hu_k^{-1}\,(a_k\,-b_k\,\hv_k)\\ [3mm]
-\om\,\hu_k\, (c_k-d_k \hv_k) ,& 1\,+\om\, \lm\, \kp_k \,\hv_k
\ea \right)\:\cdot\,({\det}_q L_k(\lm))^{-1}\,,\ee
where\\[-10mm]
\[
{\det}_q L_k(\lm)=\hv_k F_k(\lm)\,,\qquad F_k(\lm)=(b_k+\om \lm a_k\kp_k) (\lm c_k +d_k/\kp_k)\,.
\]
The expression for $L_k^{-1}(\lm)$ is singular at zeros
   $\lm'_k=-b_k/(\om a_k \kp_k)$ and $\lm''_k=-d_k/( c_k \kp_k)$ of $F_k(\lm)$.
Of course,
\be \label{Tn1Tn}
T_{n-1}(\lm)\:=\:T_n(\lm)\: L_n^{-1} (\lm)\,.
\ee
Therefore at the zeros of $F_n(\lm)$ the left-hand side is regular in $\lm$ and the
 right-hand side also has to be regular. At $\:\lm\,=\,\lm'_n\,=\,-b_n/(\om a_n \kp_n)\:$ we get
\[  A_n(\lm'_n)\: \bu_n^{-1}\: b_n/(\om \kp_n)\:+\,B_n(\lm'_n)\:=\,0\,.\]
Hence we have a formula for $\bu_n$:
\be\label{unAB}\bu_n\:=\:\lm'_n\, a_n\, B_n^{-1}(\lm'_n)\: A_n(\lm'_n)\,.\ee
{}From the condition of the regularity of the right-hand side of \r{Tn1Tn}
   at $\lm=\lm''_n=-d_n/( c_n \kp_n)$ we get
\[  A_n(\lm''_n) (-\lm''_n) \bu_n^{-1} (a_n-b_n \bv_n)+B_n(\lm''_n) (1-d_n/(\om c_n)\bv_n)=0\,.\]
Excluding $\bu_n$ by means of \r{unAB}, we obtain the formula for $\bv_n$:
\bea \lefteqn{
\bv_n\:=\,-1/(\om \kp_n)\, (A_n(\lm'_n)\,B_n(\lm''_n)\:-\,A_n(\lm''_n)\,B_n(\lm'_n))^{-1} \times}\ny\\[2mm]
\label{vnAB}&&\hs\hs\hs\times
(A_n(\lm'_n)\,B_n(\lm''_n)/\lm''_n\,-\,A_n(\lm''_n)\,B_n(\lm'_n)/\lm'_n).
\eea
Using the RTT-relations following from \r{rll}, we can permute $A_n$ and $B_n^{-1}$
in \r{unAB} to get an equivalent formula
\be\label{unABp}\bu_n\:=\:\om\, \lm'_n\, a_n\, A_n(\om\lm'_n)\: B_n^{-1}(\om\lm'_n) \,.\ee
Using \r{Alm} and \r{Be} we get \r{unPsi}.

We can also get the formulas for $\bu_{n-1}$ and $\bv_{n-1}$.
We express $L_n^{-1}(\lm)$ in terms of $A_n(\lm)$ and $B_n(\lm)$ using \r{unAB} and \r{vnAB}.
Now the formula \r{Tn1Tn} allows to find expressions for $A_{n-1}(\lm)$ and $B_{n-1}(\lm)$ in terms
of $A_n(\lm)$ and $B_n(\lm)$. Finally we substitute these expressions to \r{unAB} and \r{vnAB} in which the indices
$n$ are replaced by $n-1$. This gives us expressions for  $\bu_{n-1}$ and $\bv_{n-1}$ in
terms of $A_n(\lm)$ and $B_n(\lm)$. The described procedure can be iterated to express the local operators $\bu_k$ and $\bv_k$
in terms of $A_n(\lm)$ and $B_n(\lm)$.
For example, the result for $\bu_{n-1}$ is:
\[\fl
\bu_{n-1}=\om \lm'_{n-1} a_{n-1} \Bigl( A_n(\om\lm'_{n-1}) (\om^2\lm'_{n-1} a_n c_n+\bv_n b_n d_n/\kp_n)
 -B_n(\om\lm'_{n-1})\om \bu_n (c_n-d_n \bv_n)\Bigr)
\]
\[ \fl\hspace*{2cm}
\times\; \Bigl( -A_n(\om\lm'_{n-1}) \om\lm'_{n-1} \bu_n^{-1}(a_n- b_n \bv_n)
+B_n(\om\lm'_{n-1})(1+\om^2\lm'_{n-1}\kp_n\bv_n)\Bigr)^{-1}\,,
\]
where $\:\om\,\lm'_{n-1}\:=\:-b_{n-1}/(a_{n-1}\, \kp_{n-1})$ and the expressions \r{unABp} and \r{vnAB} for
$\bu_n$ and $\bv_n$ have to be substituted. It gives the action of $\bu_{n-1}$ on $\:|\Psi_{\bdr_n}\rangle$.
We see that the formula gets quite involved.
However, $\:\bu_1\:$ can be easily expressed in terms of $D_n$ and $B_n$:
\be \hu_1\:=\: \frac{1}{c_1}\;
D_n\left(-\frac{d_1}{c_1\kp_1}\right) B_n^{-1}\left(-\frac{d_1}{c_1\kp_1}\right)\,.\ee
For our purpose of finding matrix elements of spin operator between
eigenstates $|\Phi_{\rho,{\bf E}}\rangle$ of homogeneous $\bt_n(\lm)$
we can choose any spin operator $\bu_k$ because
they all are related by the action of translation operator having the same eigenstates $|\Phi_{\rho,{\bf E}}\rangle$.
In what follows we consider matrix elements of the spin operator $\bu_n$
because  the corresponding formula for the action \r{unPsi} is the simplest.

At the end of this section we would like to mention some similarity of our formulas
with the formulas from the paper \cite{Babelon},
where the local operators for the quantum Toda chain are expressed in terms of quantum separated variables
with the use of a recursive construction of the eigenvectors \cite{KarLeb2}.

\section{The general inhomogenous $N=2\;$ BBS-model}

In the $N=2$ case we have two charge sectors $\rho=0,\;1$. Following the language of e.g.
\cite{FZ,BL1,Lisovyy}~
the sector $\rho=0$ will be called the Neveu-Schwarz (NS)-sector, and $\rho=1$ the Ramond (R)-sector.
We are going to show that the spin matrix elements can be written in a fairly compact, although not yet factorized
form \r{mat-res},\r{Rs}. The full factorization will be achieved later for the homogenous Ising case.

\subsection{Solving the Baxter equations and norm of states}\label{BaxterNorm}

Let us fix an eigenvalue polynomial $\,t(\lm|\rho,{\bf E})\,$ of $\:\bt(\lm)\:$ corresponding to a
right eigenvector $\,|\Phi_{\rho,{\bf E}}\rangle\:$ (since in the following our chain will have the
fixed length $n$ we often shall skip the index $n$.
Also sometimes we shall suppress the arguments $\rho,{\bf E}$ in $t$).

In order to find $\,|\Phi_{\rho,{\bf E}}\rangle\:$ explicitly we have to solve
the associated $n-1$ systems ($k=1,2,\ldots,n-1$) of (right) Baxter equations:
\bea
t(-r_{n,k})\, {\tq}^{\rm R}_k(0)&=&
\left(\Delta_k^+(-r_{n,k})+\Delta_k^-(r_{n,k})\right){\tq}^{\rm R}_k(1),
\ny\\[2mm]
t(r_{n,k})\;\; {\tq}^{\rm R}_k(1)&=&
\left(\Delta_k^+(r_{n,k})+\Delta_k^-(-r_{n,k})\right){\tq}^{\rm R}_k(0).\label{BEq}
\eea
Since $\:t(\lm|\rho,{\bf E})\,$ is eigenvalue polynomial, the functional relation \r{reltau} ensures
the existence of non-trivial solutions to \r{BEq} with respect to the unknown variables
$\:{\tq}^{\rm R}_k(0)$ and $\:{\tq}^{\rm R}_k(1)\:$ for every $\,k\,=\,1,2,\ldots,n-1$.
In the $N=2$ case, this means that for every $k$ we have one
independent linear equation (in case of degenerate eigenvalues, possibly no equation).
In the case of generic parameters, both hand sides of each equation will be non-zero. So, fixing
$\:{\tq}^{\rm R}_k(0)\,=1\:$ we obtain two equivalent expressions for $\;{\tq}^{\rm R}_k(1)$:
\be {\tq}^{\rm R}_k(1)=\frac{t(-r_{n,k})}{\Delta_k^+(-r_{n,k})+\Delta_k^-(r_{n,k})}=
\frac{\Delta_k^+(r_{n,k})+\Delta_k^-(-r_{n,k})}{t(r_{n,k})}\,.\ee
Analogously from the left-Baxter equations, fixing
${\tq}^{\rm L}_k(0)=1$ we obtain
\[
{\tq}^{\rm L}_k(1)=\frac{(-1)^{n-1} t(-r_{n,k})}{\Delta_k^+(r_{n,k})+\Delta_k^-(-r_{n,k})}=
\frac{\Delta_k^+(-r_{n,k})+\Delta_k^-(r_{n,k})}{(-1)^{n-1}t(r_{n,k})}\,.
\]
Since for generic parameters $\:t(r_{n,k}|\rho,{\bf E})\:\neq\:0\:$ these explicit formulas give
\[
{\tq}^{\rm L}_k(\rho_{n,k})\;{\tq}^{\rm R}_k(\rho_{n,k})\:=\:
(-1)^{\rho_{n,k} (n-1)}\: t((-1)^{\rho_{n,k}}\: r_{n,k})/t(r_{n,k})\,.
\]
To get the periodic state, we have to insert the Skylanin-separation factor \r{QR}. Now for $\:N=2\;$
the functions $\:w_p\:$ are simple:
\be
w_p(0)\,=\,1\,,\hq w_p(1)=\frac{y}{1+x}=\frac{1-x}{y}\,,\hq (w_p(1))^2\,=\,\frac{1-x}{1+x}\,.
\ee
In the Sklyanin factor we have to use the Fermat point $p_{n,l}^{n,m}=(x_{n,l}^{n,m},y_{n,l}^{n,m})$
defined by the coordinate $\;x_{n,l}^{n,m}=r_{n,m}/r_{n,l}\,.\;$ Here it can be expressed it in terms
of $x_{n,l}^{n,m}$ only and we get
\be\fl
\frac{\langle \Phi_{\rho,{\bf E}}|\Phi_{\rho,{\bf E}}\rangle}{\langle \tilde\Psi_{\rho,\bdr'_n}|\tilde\Psi_{\rho,\bdr'_n}\rangle}\;
=\;\sum_{\bdr'_n}
\frac{{\prod_{l<m}^{n-1}(-1)^{\rho_{n,l}+\rho_{n,m}} (r_{n,m}+r_{n,l})^2}\;
\:\prod_{k=1}^{n-1}{\tq}^{\rm L}_k(\rho_{n,k}){\tq}^{\rm R}_k(\rho_{n,k})}
{\prod_{l<m}^{n-1} \;((-1)^{\rho_{n,l}} r_{n,l}+
(-1)^{\rho_{n,m}}r_{n,m})^2}\,.\ee
We can normalize to a convenient
reference state. For the moment, simple formulas arise if for the normalization we chose the auxiliary state
$\;|\tilde\Psi_{0,{\bf 0}}\rangle$ where $\;{\bf 0}=(0,0,\ldots,0)$.
{}From \r{norm} we get
\be
\frac{\langle \tilde\Psi_{\rho,\bdr'_n}|\tilde\Psi_{\rho,\bdr'_n}\rangle}
{\langle \tilde\Psi_{0,{\bf 0}}|\tilde\Psi_{0,{\bf 0}}\rangle}=
\frac{\prod_{l<m}^{n-1} (r_{n,m} (-1)^{\rho_{n,m}}+r_{n,l}(-1)^{\rho_{n,l}})}
{\prod_{l<m}^{n-1} (r_{n,m}+r_{n,l})}\,.\label{Naux}\ee
Combining all these formulas we get for the left-right overlap of the transfer matrix eigenvectors of the periodic BBS model at $N=2$:
\be\fl
\frac{\langle \Phi_{\rho,{\bf E}}\,|\,\Phi_{\rho,{\bf E}}\rangle}{\langle \tilde\Psi_{0,{\bf 0}}|\tilde\Psi_{0,{\bf 0}}\rangle}\;=\;
\frac{{\prod_{l<m}^{n-1} (r_{n,m}+r_{n,l})}}{\prod_{l=1}^{n-1}\;t(r_{n,l})}
\;\sum_{\bdr'_n}\; \frac{\prod_{l=1}^{n-1} (-1)^{\rho_{n,l}}\:t((-1)^{\rho_{n,l}}r_{n,l})}
{\prod_{l<m}^{n-1} ((-1)^{\rho_{n,m}}r_{n,m}\,+(-1)^{\rho_{n,l}}r_{n,l})}.\label{phiphi}\ee

This formula is not yet very useful since from \r{EV} it contains the summation
over the $n-1$ $\;{\mathbf Z}_2$-variables $\,\bdr'_n\,$ defined in \r{rhoprime}.
However, in \cite{gipst1} it is shown how to perform this sum explicitly, and the fully factorized result is
\be\label{normP}
\frac{\langle \Phi_{\rho,{\bf E}}\,|\,\Phi_{\rho,{\bf E}}\rangle}{\langle \tilde\Psi_{0,{\bf 0}}|\tilde\Psi_{0,{\bf 0}}\rangle}\;=\;
  2^{n-1}\,\tilde r_n'\;
  \frac{{\prod_{l<m}^{n-1} (r_{n,m}+r_{n,l})}}{\prod_{k=1}^{n}\prod_{l=1}^{n-1} (r_{n,l}\,+\,\mu_k)}
\;\prod_{i<j}^n (\mu_i\,+\,\mu_j)\,,\ee
where $-\mu_i$ are the zeros of the eigenvalue polynomial of $\bt(\lm|\rho,{\bf E})$:
\be    \bt(\lm|\rho,{\bf E})|\Phi_{\rho,{\bf E}}\rangle\;=\;
 \Lambda\:\prod_{i=1}^n\:(\lm\,+\,\mu_i)\,|\Phi_{\rho,{\bf E}}\rangle. \ee
We don't specify the factor $\Lambda$, since in the following it will cancel.

\subsection{Matrix elements between eigenvectors of the periodic $N=2$ BBS model}\label{NtwoVector}
In \r{unPsi} we obtained the action of $\bu_n$ on an eigenvector
$|\Psi_{\bdr_n}\rangle$ of $B_n(\la)$: the result is a linear combination of the original vector
plus a sum of vectors which each have one component of $\bdr_n$ shifted. In order to get the matrix
elements of $\bu_n$ in the periodic model, using \r{tPsi} we first pass to charge eigenstates
$\langle \tilde\Psi_{\rho,\bdr'_n}|$, $|\tilde\Psi_{\rho,\bdr'_n}\rangle$:
\be
\langle \tilde\Psi_{\rho,\bdr'_n}|=\langle \Psi_{0,\bdr'_n}| + (-)^\rho \langle \Psi_{1,\bdr'_n}|\,,\quad
|\tilde\Psi_{\rho,\bdr'_n}\rangle= | \Psi_{0,\bdr'_n}\rangle + (-)^\rho |\Psi_{1,\bdr'_n}\rangle.
\ee
Since $\om=-1$, $\bu_n$ anti-commutes with $\bV_n$ so that only matrix elements of $\bu_n$ between
states of different charge $\rho$ can be nonzero.
In the following we shall chose the right eigenvector from $\rho=1$, then
the left eigenvector must have $\rho=0$ (the opposite choice gives a different sign in \r{ufir}).
Using \r{unPsi}, we find
\be
\frac{\langle \tilde\Psi_{0,\bdr'_n}|\bu_n|\tilde\Psi_{1,\bdr'_n}\rangle}
{\langle \tilde\Psi_{0,\bdr'_n}|\tilde\Psi_{0,\bdr'_n}\rangle}\:=\:
\frac{a_n}{\tilde r_n}(-1)^{\tilde  \rho'_n}\:-\:\frac{\kp_1\kp_2\cdots\kp_{n-1} b_n}{r_{n,0}}\,,\label{ufir}
\ee
\bea\fl \frac{\langle \tilde\Psi_{0,{\bdr'}^{+k}_n}|\bu_n|\tilde\Psi_{1,\bdr'_n}\rangle}
{\langle \tilde\Psi_{0,\bdr'_n}|\tilde\Psi_{0,\bdr'_n}\rangle}
&=&\frac{\tilde r_{n-1} a_n b_n c_n}{r_n r_{n,0}}
\left(1+\frac{(-1)^{\rho_{n,k}} d_n}{\kp_n c_n r_{n,k}}\right)
\frac{(-1)^{\tilde  \rho'_n}\;\prod_{l=1}^{n-2}y^{n,k}_{n-1,l}}{\prod_{s\ne k}
  (r_{n,k}(-1)^{\rho_{n,k}}+r_{n,s}(-1)^{\rho_{n,s}})}\,.\ny\\ &&
\label{psiupsi}\eea
Of physical interest are the matrix elements between periodic eigenstates. To get these we have to form linear combinations
determined by the solutions of the Baxter equations: Recall \r{EV}: $|\Phi_{\rho,{\bf E}}\rangle \:=\:
\sum_{\bdr'_n} \; {\cal Q}^{\rm R}(\bdr'_{n}|\,\rho,{\bf E})\;\:|\tilde\Psi_{\rho,\bdr'_n}\rangle$
and the corresponding left equations.

Let $\langle \Phi_0|$ be a left eigenvector  of the transfer-matrix ${\bf t}_n(\lm)$ with $\rho=0$
and $|\Phi_1\rangle$ be a right eigenvector with $\rho=1$ (often suppressing the subscripts ${\bf E},\;{\bf E}'$):
\be \fl  \langle\Phi_{0,{\bf E}'}|\;{\bf t}(\lm|0,{\bf E}')\;\;=\;t^{(0)}(\lm)\,\langle\Phi_{0,{\bf E}}'|\,,\hs
  {\bf t}(\lm|1,{\bf E})\;|\Phi_{1,{\bf E}}\rangle\;=\;t^{(1)}(\lm)\,|\Phi_{1,{\bf E}}\rangle. \ee
Let $\tq^{L(0)}_k(\rho_{n,k})$ and $\tq^{R(1)}_k(\rho_{n,k})$ be the solutions of Baxter equation
  corresponding to these two eigenvectors.
After some simplification we get for the matrix elements (keeping the normalization by the auxiliary ``reference''
 state):

\be\label{mat-el}\fl
\frac{\langle\, \Phi_0\,|\,\!\sz{n}\,\!|\,\Phi_1\,\rangle}
{\langle \,\tilde\Psi_{0,\bnul}\,|\,\tilde\Psi_{0,\bnul}\,\rangle}
\,=\,\sum_{\bdr'} \CN(\bdr')
\left(R_0(\bdr') \left(\frac{a_n}{\tilde r}(-1)^{\tilde  \rho'}-
\frac{\kp_1\kp_2\cdots\kp_{n-1} b_n}{r_{0}}\right) \!
+\sum_{k=1}^{n-1} R_k(\bdr') \right),
\ee
where
\bea\label{NoN}\fl
\lefteqn{\CN(\bdr')= (-1)^{n\tilde \rho'}\, \prod_{l<m}^{n-1}\:
\frac{r_{l}\:+\,r_{m}}{r_l(-1)^{\rho_l}\:+ \,r_m(-1)^{\rho_m}},
\hx\;
R_0(\bdr')=\prod_{l=1}^{n-1}{\qs}^{{\rm L}(0)}_l(\rho_{l}){\qs}^{{\rm R}(1)}_l(\rho_{l}),}\label{NRn}\\[-1mm]
&& R_k(\bdr')\:=\:-\,\frac{a_n b_n c_n}{r_{0}}\;
Q^{\rL(0)}_k(\rho_{k}+1)\:Q^{\rR(1)}_k(\rho_{k})\prod_{l\ne k}^{n-1} Q^{\rL(0)}_l(\rho_{l})\:Q^{\rR(1)}_l(\rho_{l})\;\times
\ny\\[-1mm]  && \hs\hs\times\;
\lk 1-\frac{d_n}{\kp_n c_n\, \mun_{k}}\rk
\frac{\mun_k^{n-1}\chi_k }{\prod_{s\ne k} (\mun_{k}\,-\,\mun_{s})}\,.\label{Rk}
\eea   with $r_k=r_{n,k}$, $\rho_k=\rho_{n,k}$, $k=0,1,\ldots,n-1$,
\beq\mun_k=-r_k(-1)^{\rho_k}\,,\hx \;\tilde r=r_{0}\,r_{1}\cdots r_{n-1}\,\hx \;\;\mbox{and}\;\hx
\;\tilde\rho'={\textstyle\sum_{k=1}^{n-1}}\:\rho_k.\eeq
The origin of the different terms in \r{mat-el} is: the sum over $\bdr'$ comes from \r{EV},
$\CN(\bdr')$ is the normalization factor from \r{Naux}.
The terms at $R_0(\bdr')$ arise from the first line of \r{unPsi}: the shift in $\rho_{n,0}$ affects the
charge sector only. The sum over $k$ and expression for $R_k(\bdr')$
come from the second line in \r{unPsi}. Now, the sum over $k$ can be performed. Indeed,
as shown in \cite{gipst2}, using the Baxter equations, some cancellations take place
and \r{mat-el} can be written as
\be \frac{\langle\,\Phi_0\,|\,\bu_n\,|\,\Phi_1\,\rangle}
{\langle\, \tilde\Psi_{0,{\bf 0}}\,|\,\tilde\Psi_{0,{\bf 0}}\,\rangle}
\;=\;\frac{a_n}{2\,r_0}\sum_{\bdr'\in{\mathbb Z}_2^{n-1}} \CN(\bdr')\,R_0(\bdr')\:R(\bdr')\label{mat-res}\ee
with\\[-7mm]
\be\fl R(\bdr')\;=\;\frac{t^\nul(-\fba_n)}
    {\prod_{l=1}^{n-1}(-\fba_n+(-1)^{\rho_l}r_l)}\:+\:
\frac{t^\one(\fba_n)}{\prod_{l=1}^{n-1}(\fba_n+(-1)^{\rho_l}r_l)},\hs\hx \fba_n\,=\,\frac{b_n}{a_n\,\kp_n}\,.\label{Rs}\ee

Despite the simple appearance, for the general inhomogenous $N=2$ BBS-model, performing the sums
over the ${\mathbb Z}_2$ variables explicitly seems to be a presently hopeless task.
However, for the homogenous Ising model we shall show this to be possible.


\section{Homogeneous $N=2\;$ BBS-model}

\subsection{Spectra and zeros of the $B_n$- and $t_n$-eigenvalue polynomials}

We now specialize to $N=2$ taking all parameters site-independent (``homogenous''):
\be \fl a_m=a,\;\;b_m=b,\;\;c_m=c,\;\;d_m=d,\;\;\kp_m=\kp,\;\;r_m=r,\;\;
{\cal L}_m(\lm^2)={\cal L}(\lm^2),\hx\forall\:m.\label{hom}\ee Then the classical monodromy is\\[-5mm]
\be\label{calABCDhom}
\left(\ba{cc}
{\cal A}_n(\lm^2) &  {\cal B}_n(\lm^2)\\
{\cal C}_n(\lm^2) &  {\cal D}_n(\lm^2)
\ea\right)=\left({\cal L}(\lm^N)\right)^n\!.\ee
Consider trace, determinant and eigenvalues $x_\pm$ of $\;{\cal L}$:
\be\label{taulm}
\tau(\lm^2)=\tr {\cal L}(\lm^2)=
1+\frac{b^2 d^2}{\kp^2} - \lm^2 (\kp^2 + a^2 c^2), \ee
\be\label{deltalm} \delta(\lm^2)= \det\, {\cal L}(\lm^2)\, =\,
(b^2/\kp^2-\lm^2a^2)\,(d^2-\lm^2c^2\kp^2)\,=\,F(\lm)\,F(-\lm), \ee
\be x_{\pm}={\textstyle\frac{1}{2}}(\tau\:\pm \,\sqrt{\tau^2\,-\,4\,\delta}),\hq
   F(\lm)\:=\:(b-a\kp\lm)(\lm c+d/\kp).\label{xpm}  \ee
{}From the matrix ${\cal L}(\lm^2)$ we obtain
\be {\cal B}_m(\lm^2)\:=\:-\lm^2 \,(a^2\,-b^2)\,(x_+^n\,-\,x_-^n)/(x_+\,-x_-),\ee
so that the zeros of $\;{\cal B}_m$ are at $\;x_+/x_-\:=\:\e^{{\rm i}\,m\,\phi_{n,s}}$ with
\be
\label{zerphi} \phi_{n,s}=2\pi s/n, \qquad s=1,2,\ldots,n-1,\quad s\neq 0.\ee
Using $\;\tau^2\,=\,4\,\delta\;{\cos}^2(\phi/2)\:$ and \r{taulm}, \r{deltalm}
we can translate the zeros labeled by $\,\phi_{n,s}\,$ by a quadratic equation
in $\,\lm^2\:$ into zeros $\lm_{n,s}$.

Now we solve the functional relations \r{rectau},\r{reltau} for the transfer matrix spectrum.
Using \r{rectau} for $j=2$ and eliminating $\tau^{(3)}$ by \r{reltau}
we get the functional relation
\be\label{frN2}
t(\lm)\;t(-\lm)\:= (-1)^\rho (z(\lm)+z(-\lm))+{\cal A}_n(\lm^2)+{\cal D}_n(\lm^2)
\ee  which we shall use to find $t(\lm)$.
In terms of \r{taulm} and \r{deltalm} this reads
\be  t(\lm)\;t(-\lm)\:= (-1)^\rho\left(\delta_+^n+\delta_-^n\right)\,+\:x_+^n\:+\,x_-^n. \ee
where $\;\delta_\pm\:=\:(b\:\pm\, a\kp\lm)\;(d\:\mp\, c\kp\lm);\hx \delta_+\delta_-\:=\:\delta(\lm^2)\:=\:x_+\,x_-.$
Introducing $\,\qu\,$ taking the $\,n\,$ values $\;\pi(2s+1-\rho)/n,\hx s\,=\,0,\ldots,n-1\,$, we can write \r{frN2}
as
\beq\fl
t(\lm)\:t(-\lm)= (-1)^n\prod_{\qu}(e^{{\rm i}\qu} \delta_+\!-\tau(\lm^2)+
e^{-{\rm i}\qu} \delta_-)=(-1)^n \prod_\qu
\lk A(\qu)\lm^2\!-C(\qu)+2{\rm i}\,B(\qu)\lm\rk\eeq
with \\[-12mm]  \bea  \lefteqn{ A(\qu)\,=\,a^2\,c^2\,-2\kp\, ac\,\cos \qu\,+\kp^2;\hs
B(\qu)\,=\,(ad-bc)\sin \qu\,;}\ny\\ [2mm] &&
\hs\hs C(\qu)\,=\,1\,-\,2(b\,d/\kp)\,\cos \qu \,+\, b^2\,d^2/\kp^2\,.\eea
Factorizing the polynomial in $\lm$ we get
\be t(\lm)\;t(-\lm)\:=\:(-1)^n\prod_\qu A(\qu)\:(\la\,-\,\lms_\qu)\,(\la\,+\,\lms_{-\qu}) \label{abc} \ee
with   \\ [-12mm]
\be  \lms_\qu\,=\,\frac{1}{A(\qu)}(\sqrt{D(\qu)}\,-{\rm i} B(\qu)),\qquad
D(\qu)\;=\;A(\qu)\,C(\qu)\,-B(\qu)^2,        \label{acb}\ee
(fixing the sign of $\sqrt{D(q)}$ requires a special convention, see \cite{gips})
and after some arguments we find the spectrum
\be\label{tNSR}
t(\lm)\;=\;(a^n c^n+(-1)^\rho\kp^n)\;\:{\textstyle\prod_{\qu}}\: (\lm\,\pm\,\lms_\qu),
\ee
where the signs are not yet fixed. Comparing
the $\lm$-independent term in \r{tE}
\be  \label{tlm}\fl
t(\lm)\:=\:1\:+(-1)^\rho\:b^n\, d^n/\kp^n\,+E_1\lm+\cdots+E_{n-1}\lm^{n-1}
+\lm^{n} (a^n c^n+(-1)^\rho \kp^n). \ee
with the corresponding term in (\ref{tNSR}) shows that the number of minus
signs in \r{tNSR} must be even (odd) for the NS-sector $\rho=0$ (R-sector $\rho=1$).

It is useful to introduce the following notion: The eigenvalue \r{tNSR} with all
$+$-signs is called to possess ``no quasi-particle'' excitations. Each factor labeled by $\qu$
having a minus sign is said to contribute the ``excitation of the $\qu$-quasi-momentum''.
We shall accordingly label the minus signs by a set of variables $\sig_\qu\:\in\:{\mathbb Z}_2$, where
for unexcited (excited) levels $\qu$ we put $\sig_\qu\,=0$ ($\sig_\qu\,=1$). So instead of \r{tNSR}, we shall write
more precisely
\be\label{tZ}
t^{(\rho)}(\lm)\;=\;(a^n c^n+(-1)^\rho\kp^n)\;\:{\textstyle\prod_{\qu}}\: (\lm\,+(-1)^{\sig_\qu}\,\lms_\qu).
\ee

The corresponding eigenvectors have been considered for the inhomogenous case
in Subsection \ref{NtwoVector}.

\subsection{Functional relation for the diagonal-to-diagonal Ising model transfer-matrix}\label{fct}

In this subsection we specialize the results of the previous subsection to
the case of the diagonal-to-diagonal transfer-matrix
of the Ising model on a square lattice \r{ISIdiag}.
So, we set $a=c$, $b=-d$, $\kp=1$ and $\lm=b/a$.
Let us calculate the ingredients of the functional relation \r{frN2}. We have
$\,F_m(\lm)\:=\:-(b-a \lm)^2$. Therefore due to \r{zDelta},
$z(\lm)\:=\:(-1)^n (b\,+a\,\lm)^{2n}$, $\:z(b/a)\:=\:(-1)^n\: (2b)^{2n}$, $\:z(-b/a)=\,0\:$
and the averaged $L$-operator \r{avL} at $\lm^2=b^2/a^2$ becomes
\[\fl
{\cal L}_k(b^2/a^2)\:=\:\left( \ba{cc}
1-b^2/a^2,\ &  -b^2/a^2\,(a^2-b^2)\\ [3mm]
a^2-b^2 ,& b^2 (b^2 -a^2)
\ea \right)\,= \left( \ba{l}\! 1\!\\[2mm]\! a^2\!\!
\ea \right)\cdot (1-b^2/a^2)\cdot \left( \ba{ll}\!\!
1,\ &\!\!\!  -b^2\!\!
\ea \right)\,.
\]
Hence
\[
{\cal A}_n(b^2/a^2)+{\cal D}_n(b^2/a^2)\;=\;\mathrm{tr}\; {\cal T}_n(b^2/a^2)\;=\;(1-b^2/a^2)^n (1-a^2 b^2)^n\,.
\]
Substituting these expressions into \r{frN2}, we get the following functional relation
\[
t(b/a)\;t(-b/a)\:=\:(-1)^{\rho+n} (2b)^{2n}+(1-b^2/a^2)^n (1-a^2 b^2)^n\,.
\]
We want to compare this with the functional relation equation (7.5.5) in \cite{Baxter:book}:
\[
V(K,L)\; V(L+{\rm i}\pi/2,-K)\; C\:=\:(2{\rm i} \sinh 2L)^n \:I+ (-2{\rm i} \sinh 2K)^n \:R\,,
\]
where $C$ is the operator of translation, $R$ is the operator of spin flip ${\bf V}_n$
and $V(K,L)$ is the transfer-matrix \r{t_triangle} with
$K_x=0$, $K_y=L$, $K_d=K$, $e^{-2L}\,=\,{b}/{a}$, $\tanh K=a\,b$.
Therefore $V(K,L)=\exp(n L)\:\cosh^n K\;{\bf t}_n(b/a)$.
Similar analysis gives $V(L+{\rm i}\pi/2,-K)C\,=\,{\rm i}^n \exp(n\, L)\:\cosh^n\!K\;{\bf t}_n(-b/a)$.
Now taking into account that the eigenvalues of $\,R\:$ are $(-1)^\rho$ and
\[\fl
2\sinh 2K\;=\;\frac{4ab}{1-a^2b^2},\qquad  2\sinh 2L\;=\;\frac{a^2-b^2}{ab},\qquad
\frac{\exp(-2L)}{\cosh^2 K}\;=\;(1-a^2b^2)\, b/a\,,
\]
we see that both functional relations are identical.

\subsection{Ising model: Spectra and zeros of the $B_n(\lm)$- and $t_n(\lm)$-eigenvalue polynomials}

We now specialize further to the Ising case \r{ISI} as already advertised in Subsection \ref{isin}:
\be a_j\:=\:c_j\;=\;a,\hs
b_j\:=\:d_j\;=\;b,\hs \kp_j\,=\,1;\hs\forall\:j\,.  \label{isi}\ee
In the Ising case \r{isi} the $2^n$ eigenvalues of \r{tZ} with \r{acb} can be written ($2^{n-1}$ in each
sector $\rho=0,\,1$):
\be\label{tmgen}\fl
t^{(\rho)}(\lm)=(a^{2n}\!+(-1)^\rho)\,\prod_{\qu}\,(\lm\,+(-1)^{\sig_\qu}\,\lms_{\qu}),\hx\;\;
\lms_{\qu}\,=\,\lms_{-\qu}\,=\,\sqrt{\frac{b^4\,-2\, b^2\cos\qu+1}{a^4\,-2\, a^2\cos\qu+1}}\,,
\ee
where the quasi-momentum $\qu$ in each sector takes $n$ values:
\be \fl\mbox{$\hs\qu={\displaystyle\frac{2\pi}{n}\,m},\hs $ $m$ integer for $\rho=1$ (R); ~~$m$
half-integer for $\rho=0$ (NS).}\label{mmm}\ee
Recall that we found from \r{tlm} that in the NS (R) sector, the eigenstates of ${\bf t}(\lm)$
have an even (odd) number of excitations: $\prod_{\qu}\,(-1)^{\sig_\qu}\,=\,(-1)^\rho$.

For $\qu=0$ (this occurs for R-sector only) and $\qu=\pi$  we define
\be  \lms_{0}=\frac{b^2-1}{a^2-1}\,,\quad \hs\lms_{\pi}=\frac{b^2+1}{a^2+1}\,.\label{s}\ee
$\qu=\pi$ is in the R sector for $n$ even. However, for $n$ odd it is in the NS sector.
The different presence of factors $(\lm\,\pm \lms_0)$ and $(\lm\,\pm\lms_\pi)$ in \r{tmgen} for $n$
even or odd often makes it necessary to consider the cases $n$-even and $n$-odd separately.
In the following we shall reserve the notation $\lm_{\qu}$ for $\lm_{\qu}=(-1)^{\sigma_{\qu}}\lms_{\qu}\;$ and
otherwise use $s_\qu$ as defined in \r{tmgen}.

The zeros $\lm_{n,k}$ of the $B_n(\lm)$ eigenvalue polynomial are determined by \r{zerphi},\r{taulm},\r{deltalm}:
\be\fl \hq\hq \tau(\lm_{n,k}^2)\;=\;4\cos^2{q_{n,k}}\,F(\lm_{n,k})\:F(-\lm_{n,k}),
               \hq q_{n,k}=\,\pi\,k/n,\hq k=1,\ldots,n-1.  \label{kkk}\ee
Since now\\[-10mm]
\be F(\lm)\,=\,F(-\lm)\,=\,b^2\,-\,a^2\,\lm^2;\hq \tau(\lm^2)\,=\,1\,+\,b^4\,-(1\,+\,a^4)\,\lm^2,  \ee
we get\\[-7mm]
\be  r_{n,k}\:=\:\sqrt{(b^4\,-2\, b^2\cos{q_{n,k}}+1)/(a^4\,-2\, a^2\cos{q_{n,k}}+1)}=\lms_{q_{n,k}} \,. \ee
Observe that $\:\lms_{\qu}\:$ and $\:r_{n,k}$ may coincide.

\subsection{Ising model state vectors from Baxter equations}\label{zweizwei}

In order to obtain the eigenvectors of ${\bf t}(\lm)$, we have to solve Baxter's
equations. For our restricted parameters \r{isi} we have  $\:F(\la)=F(-\la)\,$ and
the left and right Baxter equations \r{BAXL}, \r{BAXR} become identical.
Omitting the superscripts $L$ and $R$ on $Q_k$ and recalling
$\;\la_{n,k}=-(-1)^{\rho_{n,k}}r_{n,k}, \;\;\rho_{n,k}=0,1\:$ we obtain:
\be \fl t_n(\la_{n,k})\,Q_k(\rho_{n,k}) \,=\,\lk\frac{(-1)^\rho F^{n-1}(\la_{n,k})}{(\la_{n,k})^{n-1}\,\chi_k}
\:+\:(-\la_{n,k})^{n-1}\,\chi_k\:F(\la_{n,k})\:
     \rk Q_k(\rho_{n,k}+1).  \label{BaxIs}\ee
{}From \r{BaxIs} we get the following compatibility condition:
\[
t(-r_{n,k}) t(r_{n,k}) =(-1)^{n-1 }
\left(\frac{(-1)^\rho\:F^{n-1} (r_{n,k})}{(r_{n,k})^{n-1}\,\chi_k}
\,+\,(-r_{n,k})^{n-1}\,\chi_k\,F(r_{n,k})\right)^2\!\!,
\]
if $t(\lm)$ is an eigenvalue from the sector $\rho$.
If $\,(-1)^{k}=(-1)^{\rho+1}$ then the quasi-momentum $\qu=q_{n,k}$ belongs to the sector $\rho$
and for $r_{n,k}=\lms_{q_{n,k}}$ we have $t(-r_{n,k})\,t(r_{n,k})=0$. This implies
a relation not depending on a particular $t(\lm)$ and its $\rho$:
\be\label{shchi}
\chi_k^2\: r_{n,k}^{2(n-1)}=(-1)^{n+k+1}F^{n-2}(r_{n,k})\,.
\ee
Although the eigenvalue polynomial $t(\lm)$ is known from \r{tmgen}, to solve \r{BaxIs} for
the $Q_k(\rho_{n,k})$ can meet a difficulty if $t_n(\la_{n,k})$ vanishes or if, due to \r{shchi},
the big bracket on the right of \r{BaxIs} vanishes. All this can happen and we have to distinguish
four cases
(we suppress $n$ and write just $r_k\,=\,r_{n,k}$ and $\rho_k\,=\,\rho_{n,k}$):\\[3mm]
\noindent
(i)$\hq$ \underline{$(-1)^\rho=(-1)^{k}$:}$\hs$ This is the easy case, since from \r{mmm} and \r{kkk}\\[2mm]
\hq\; $\,t^{\rho}(r_k)\ne 0$ and $t^{\rho}(-r_k)\ne 0$, and we may normalize and solve
\[Q_k^{\rm L,R}(0)\,=\,1\,,\hq
Q_k^{\rm L,R}(1)\:=\:\frac{(-1)^{n-1}t^{\rho}(-r_k)}{2 \chi_k\:r_k^{n-1}\:F(r_k)}\,.
\]
The other three cases occur for
\underline{$(-1)^\rho=(-1)^{k-1}:$}$\hs$
\\[1mm]
(ii)\hq
$t^{\rho}(r_k)\ne 0,\ t^{\rho}(-r_k)=0$:
$t^{\rho}(\lm)$ contains a factor $(\lm+r_k)^2$
(both $\qu=\pm q_k$ not excited), we may normalize
\[Q^{\rm L,R}_k(0)=1, \hq Q^{\rm L,R}_k(1)\:=\:0\,.\]
(iii)\hq $t^{\rho}(r_k)=0,\ t^{\rho}(-r_k)\ne 0$:
$t^{\rho}(\lm)$ contains a factor $(\lm-r_k)^2$ (both $\qu=\pm q_k$ are excited),
       we cannot choose $Q^{\rm L,R}_k(0)\,=\,1$, but we may normalize
       \[Q^{\rm L,R}_k(0)\:=\:0\,,\hq Q^{\rm L,R}_k(1)\:=\:1\,.\]
(iv)\hq
$t^{\rho}(r_k)=t^{\rho}(-r_k)=0$: $t^{\rho}(\lm)$
contains $(\lm^2-r_k^2)$
(either $\qu=+q_k$ or $\qu=-q_k$ is excited): A L'H{\^o}pital procedure, using a slight
perturbation of \r{isi} as described in \cite{gipst1},
is required (to obtain eigenvectors of translation operator), leading to
\[
Q^{\rm R}_k(0)=Q^{\rm L}_k(0)=1\,,\hx
Q^{\rm R}_k(1)=-Q^{\rm L}_k(1)=
\frac{(-1)^{n+\sig_{q_k}+1}  2{\rm i}\sin{q_k}\,{t}^{\rho}_{\check q_k}(-r_k)}
{ n\, \chi_k\:r_k^{n-1}\: A(q_k)}\,\]
(observe that from the L'H{\^o}pital-limit $\;Q^{\rm R}_k(1)\,=\,-\,Q^{\rm L}_k(1))$, where
\be \fl
t^{\rho}(\lm)\:=\:{ t}^{\rho}_{\check q_k}(\lm)\;
(\lm+(-1)^{\sig_{q_k}}\lms_{q_k})(\lm-(-1)^{\sig_{q_k}}\lms_{-q_k}),
\hq A(\qu)\:=\:a^4-2a^2\cos{\qu}+1. \label{tcheck}\ee
In the following we shall consider only the three cases which allow the normalization $Q^{\rm L,R}_k(0)=1$.
Case (iii) can be treated too, but requires a special treatment, which here we shall not enter.
According to which case the corresponding eigenvalue polynomial belongs, let us define
the sets $\CDC^{(\rho)}$, $\WD^{(\rho)}$, $\CD^{(\rho)}$:\\[0.5mm]
$\hspace*{27mm} k\in\CDC^{(\rho)}\;\:$ if $t^{\rho}$ has a factor $(\lm+r_k)^2$, i.e. we have case (ii),\\
$\hspace*{27mm} k\in\WD^{(\rho)}\;\;$ if $t^{\rho}$ has a factor $(\lm-r_k)^2$, case (iii), and\\
$\hspace*{27mm} k\in\CD^{(\rho)}\;\;$ if $t^{\rho}$ has a factor $(\lm^2-r_k^2)$, i.e. we have case (iv).\\[1mm]
By $D=|\CD|$ we denote the number of elements in $\CD=\CD^{(0)}\cup \CD^{(1)}$, similarly for $\CDC$, etc.

\section{Calculation of the matrix elements of $\sig_n^z$ in the homogeneous Ising model}

\subsection{Explicit evaluation of the factors $\;\CN(\bdr')\:R_0(\bdr')\:R(\bdr')\,$ in \r{mat-res}}

We now start to evaluate \r{mat-res} with \r{NoN} and \r{Rs} for the homogenous Ising model where
the parameters simplify drastically. Now
\be \fba\:=\:b/a,\hs r^2_{0}\,=\;(a^2-b^2)(a^{4n}-1)/(a^4-1).\label{avier}\ee and
$\bu_n$ is represented by the Pauli $\sig_z$.

We had agreed to consider initial states from the R-sector. Then for matrix elements of $\sig_z$ the
final state must be NS. We specify the initial state by the momenta which are excited, i.e. by the
$\sig_k$ which are one, analogously the final state. Excluding for the time being case (iii), we take $\WD^{(\rho)}$ to be
empty.

On the right of \r{NoN} we have to evaluate the factors $\;\CN(\bdr')\:R_0(\bdr')\:R(\bdr')\,$. Let us
start with
$\;\;R_0(\bdr')\:=\:\prod_{l=1}^{n-1}{\qs}^{(0)}_l(\rho_{l}){\qs}^{(1)}_l(\rho_{l})\,.$

For any choice of excitations, always one of the factors $\qs^{(0)}_l(\rho_{l})$ or $\qs^{(0)}_l(\rho_{l})$ is from case
(i) of Subsection \r{zweizwei}. Since we exclude for the moment case (iii), the other factor then must be from (ii) or (iv).
So always $Q_l^{(0)}(0)Q_l^{(1)}(0)\,=\,1$.
For $l\in \CDC$, case (ii), we have $\;\qs^{(0)}_l(1){\qs}^{(1)}_l(1)=0\;$ since either
$\qs^{(0)}_l(1)=0\;$ or $\;{\qs}^{(1)}_l(1)=0\;$ depending on the parity of $l$.
So, in (\ref{NRn}) the summation reduces to the summation over $\rho_l$ for $l\in \CD\;$ only, with
fixed $\;\rho_l=0\;$ for $\;l\in \CDC$.

$R_0(\bdr')\:$ receives non-trivial contributions from $\:Q_k(1)\:$ of cases (i) and (iv). However,
these can be written in a simple way if we use the explicit formulae for $\:t^{(\rho)}(-r_k)\,.$ For both
values $\rho_l=0,\:1$ the result is
\be  Q^\nul_l(\rho_l)\:Q^\one_l(\rho_l)\;=\; (-1)^{(n-1)\rho_l}\,\frac{(-1)^{\rho_l}r_{l}+ \xi_l}
{r_{l}\,+\,\xi_l}\cdot\prod_{k \in \CDC}\frac{(-1)^{\rho_l}r_{l}+r_{k}}{r_{l}+r_{k}}
  \,,\label{Rnu} \ee
where we get different results according to whether $\:\lms_0\:$ or $\:\lms_\pi\:$ or both \r{s} are excited:
\be\fl \xi_l=\lb\,{(-1)^{\sigma_0}\;\:{\displaystyle\frac{b^2 - e^{{\rm i}q}}{a^2 -e^{{\rm i}q}}\;\;}\atop
    (-1)^{\sigma_0}\;\:{\displaystyle\frac{b^2\, e^{{\rm i}q}-1}{a^2 -e^{{\rm i}q}}}}
    \hq \mbox{for}
  \hx (-1)^{\sig_0}\,=\,\pm(-1)^{\sig_\pi}\,;\right.\quad
\tilde q_l=(-1)^{\sigma_{q_l}+|{\cal D}|+l}\,q_l\,.\label{xi}\ee
Now,~ multiplying by $\:\CN(\bdr')\,,$ it is easy to see that the products $k\in\CDC$ in \r{Rnu} cancel
(recall that $\rho_k=0$ for $\,k\in\CDC\,)$
and we get finally
\be \fl\CN(\bdr)\cdot R_0(\bdr')\,=\,\;\prod_{l\in\CD}(-1)^{\rho_l}\,\frac{(-1)^{\rho_l}r_{l}\,+\,\xi_l}
{r_{l}\,+\,\xi_l} \prod_{m\in\CD, m>l}\;
\frac{r_{l}+r_{m}}{(-1)^{\rho_{l}} r_{l}+ (-1)^{\rho_{m}}r_{m}}\,.\label{NRN}\ee

In the calculation of $R(\bdr')$ in \r{Rs} we have to insert our explicit expressions for
 $t^\nul(-\fba_n)$ and $t^\one(\fba_n)$ from \r{tmgen}. Here, as already mentioned after \r{s},
 the cases of even $n$ and odd $n$
 give different formulas. E.g. the factor $(\lm\,-(-1)^{\sig_\pi}\,\lms_\pi)$
 is present only for R $n$ even and NS $n$ odd. So
\be \fl \mbox{NS},\;\;n\;\mbox{odd:}\hx t^{(0)}(-\fba)\:=\:(a^{2n}+1)
         (-\fba\,+(-1)^{\sigma_\pi}s_\pi) \prod_{k \in \CDC^{(0)}}
               (-\fba\,+r_{k})^2 \prod_{l\in\CD^{(0)}} (\fba^2\,-r_{l}^2)\,,\label{tNS}\ee
(for even $n$ omit the bracket with $s_\pi$), since in the NS-sector
only odd $k$ appear, and these fall into one of the classes (ii) and (iv), class (iii) being
momentarily excluded. Analogously:
\be
\fl\mbox{R, $\;\;n$ odd:}\hq
t^{(1)}(\fba)\:=\:(a^{2n}-1) (\fba\,+(-1)^{\sigma_0}s_0)\!\!\prod_{k \in \CDC^{(1)}} (\fba\,+r_{k})^2
\prod_{l\in \CD^{(1)}} (\fba^2\,-r_{l}^2).
\label{tR}\ee
By slight manipulation we can move the $\rho_l$-dependent terms such that they appear
only in one place each in the numerator and get
\bea\fl \lefteqn{R^{(n\,{\rm odd})}(\bdr')\:=\:\RRR\:\cdot\:\lb (-1)^{\sig_\pi} (a^2+1)
\lk -\fba\,\,+(-1)^{\sig_\pi}\lms_\pi\rk\;
{\textstyle \prod_{l\in \CD}}((-1)^{\rho_l}\,r_l\,+\fba)\right.} \ny\\[2mm]
\label{rRo}\fl
&&\left.\hq\;-(-1)^{\sig_0} (a^2-1) \lk \fba\,+(-1)^{\sig_0}\lms_0\rk\;
{\textstyle \prod_{l\in \CD}}((-1)^{\rho_l}\,r_l\,-\fba)\rb\,\cdot\,{\textstyle \prod_{k\in \CDC}}\,((-1)^k\fba\,+r_k)
\eea
with
\be\label{RRR}\fl
\RRR\:=\:\lk \aab\abb\rk^{-(n-1)/2}a^{n-1}(a^{4n}-1)/(a^4-1),\hq \aab\,=\,a^2\,-b^2, \hq\abb\,=\,1\,-a^2\,b^2.\ee
The first term in the curly bracket comes from the NS-sector final state, the second from the R initial state.
The formula for $R^{(n\,{\rm even})}$ is similar.

\subsection{Summation, square of the matrix element}

Combining \r{NRN} with \r{rRo} the spin matrix element is given by a multiple sum over the components of $\bdr'$:
\bea\fl
\frac{\langle\, \Phi_0\,|\,\!\sz{n}\,\!|\,\Phi_1\,\rangle}
{\langle \,\tilde\Psi_{{\bf 0}}\,|\,\tilde\Psi_{{\bf 0}}\,\rangle}\;
 &=&\sum_{\bdr'\,\in\,{\mathbb Z}_2^{n-1}} \lk {\cal R}_+^\nu\;
   \prod_{l\in \CD}((-1)^{\rho_l}\,r_l\,+\fba) + {\cal R}_-^\nu\;
   \prod_{l\in \CD}((-1)^{\rho_l}\,r_l\,-\fba) \rk\;\times\ny\\
 \fl &&\times\;\prod_{l\in\CD}(-1)^{\rho_l}\,\frac{(-1)^{\rho_l}r_{l}\,+\,\xi_l}
{r_{l}\,+\,\xi_l} \prod_{m\in\CD, m>l}\;
\frac{r_{l}+r_{m}}{(-1)^{\rho_{l}} r_{l}+ (-1)^{\rho_{m}}r_{m}}\,,
  \label{matc1} \eea
with some $\bdr'$-independent factors ${\cal R}_\pm^\nu$. The superscript $\nu$ is there to remind us that
we have different expressions for $n$ even and $n$ odd, respectively. Now, in \cite{gipst2} it is shown that
this sum can be performed, resulting in a factorized expression. As an example here we quote the summation formula
for the multiple summation over $\rho_l$ with $l\in\CD$ if the dimension of $\CD$ is odd and $\:\xi_l\:$
defined by the upper formula of \r{xi}:\\[-2mm]
\bea\fl\lefteqn{\sum_{\rho_{l},\,\,l\in\CD}\; \frac{\prod_{l\in\CD}\; (-1)^{\rho_{l}}\;
     (\,(-1)^{\rho_{l}} r_{l}\,+\, \xi_l)
\;\:(\,(-1)^{\rho_{l}} r_{l}\, +\,\fba)} {\prod_{l<m,\: l,m\in\CD}\;
     (\,(-1)^{\rho_{l}} r_{l}\:+\, (-1)^{\rho_{m}}r_{m})}}\ny\\[2mm]
   \fl  &&
  =\; {\cal C}\,(b\pm a)\lk {\textstyle\prod_{j\in\CD}}\qq{j}\mp a b\rk\:
    {\displaystyle \frac{\prod_{l\in\CD}\: (2\,r_l/a)
  \:(a^2\,\qq{l}-1)^{(D-1)/2}\;(\qq{l}-a^2)^{(D-3)/2}}{{\displaystyle\prod_{l,m\in\CD,\,l<m}\;(\pm (\qqq{l}{m}-1))}}}\,,
\label{summa}\eea
\[  {\cal C}\:=\:\:\aab^{-(D-1)(D-3)/4}\:\abb^{-(D-1)^2/4}.
\]
The case of $\CD$ even is similar, see (52) of \cite{gipst2}.

In the following, we shall be interested
in the product of the matrix elements of the spin operator between arbitrary periodic states, which does not
depend on normalization of the left and right eigenstates,
i.e. we want to calculate
\be\label{ratio}
\frac{\langle \Phi_0|\bu_n|\Phi_1\rangle
\langle \Phi_1|\bu_n|\Phi_0\rangle}
{\langle \Phi_0|\Phi_0\rangle\langle \Phi_1|\Phi_1\rangle}\,.
\ee
Taking the absolute squares, several factors in \r{summa} can be re-written, e.g.
\be  |a^2\,\qq{l}\,-\,1\,|^2\:=\:|\qq{l}-a^2|^2\:=\:A({\tilde q}_l)\:=\:a^4\,-2a^2\cos{{\tilde q}_l}\,+1,
\ee
\be |\,\qqq{l}{m}\,-\,1\,|^2
   \:=\:\frac{r_m^2\,-\,r_l^2}{\aab\:\abb}\;A(\qt_m)\;A(\qt_l)\;\;
   \frac{\sin\frac{1}{2}(\tilde q_l+\tilde q_m)}{\sin\frac{1}{2}(\qt_l\,-\qt_m)}  \ee
and all factors $\aab$, $\abb$ and $A(\qt_m)$ cancel. So we get
for arbitrary $n$ and ${\sigma_0}={\sigma_\pi}$:\\[-2mm]
\bea\fl
\lefteqn{
\frac{\langle\, \Phi_0\,|\,\sz{n}\,|\,\Phi_1\,\rangle\;\langle\, \Phi_1\,|\,\sz{n}\,|\,\Phi_0\,\rangle}
{\langle \,\tilde\Psi_{0,{\bf 0}}\,|\,\tilde\Psi_{0,{\bf 0}}\,\rangle^2}
\:=\:(\lm_\pi^2\,-\,\lm_0^2\,)^{(D-\delta)/2} \:(\lm_0\,+\lm_\pi)^{\delta}\:
\prod_{l \in \CD} \frac{2\,r_l}{(\lm_0\,+r_l)\,(\lm_\pi\,+r_l)}\;\times}\ny\\
 &&\times\!\!
\prod_{l<m,\,\, l,m \in\CD}\frac{r_l\,+r_m}{r_l\,-r_{m}}\:\cdot\:
\frac{\sin\,\frac{1}{2}(\qt_l-\qt_m)}{\sin\,\frac{1}{2}(\qt_l\,+\qt_m)}
\label{ME}\eea
where $\delta=1$. In a similar way we can find the product of matrix elements in the case of
${\sigma_0}\ne {\sigma_\pi}$. The final result is \r{ME} with $\delta=0$.
Observe that the explicit appearance of excitations of
type (ii), i.e. $k\in\CDC$ has disappeared from our formula (recall that we still exclude $k\in\WD$).

\subsection{Normalization of the periodic states,
final result in terms of $\;\lm_0,\:\lm_\pi, r_k$ and ${\tilde q}_l$}

In order to compare \r{ME} to the results obtained by  A.~Bugrij and O.~Lisovyy
\cite{BL1,BL2} we change the normalization and calculate the ratio \r{ratio}.
To do this we have to divide \r{ME} by
\be\label{prq1}\langle \Phi_0|\Phi_0\rangle
\langle \Phi_1|\Phi_1\rangle/
\langle \tilde\Psi_{0,{\bf 0}}|\tilde\Psi_{0,{\bf 0}}\rangle^2\,.
\ee
However, formula \r{normP} cannot be used directly in our degenerate Ising case \r{isi}.
As in the case (iv) we have first to go off the Ising point and consider $ad-bc=\eta$ and
apply l'Hopital's rule for $\eta\rightarrow 0$. For $n$ odd the result is
\[\fl
\frac{\langle \Phi_0|\Phi_0\rangle\;\langle \Phi_1|\Phi_1\rangle}
{\langle \tilde\Psi_{0,{\bf 0}}|\tilde\Psi_{0,{\bf 0}}\rangle^2}
\;=\;2^{|{\cal D}|} \prod_{k=1}^n (2r_{n,k}) \cdot \frac{\prod_{k{\rm -odd}}(\lm_\pi\pm r_{n,k}) }
{\prod_{k \rm{-even}} (\lm_\pi+r_{n,k})}\cdot
\frac{\prod_{k \rm{-even}} (\lm_0\pm r_{n,k}) }
{\prod_{k \rm{-odd}} (\lm_0+r_{n,k})}\;\times
\]
\[\fl\hq\times\;
\frac{ \prod_{k<l, k,l{\rm -odd}} \Bigl((r_{n,k}+r_{n,l}) (\pm r_{n,k}\pm r_{n,l})\Bigr)
\prod_{k<l, k,l{\rm -even}} \Bigl((r_{n,k}+r_{n,l}) (\pm r_{n,k}\pm r_{n,l})\Bigr)}
{\prod_{k{\rm -odd},l{\rm -even}}  \Bigl((\pm r_{n,k}+r_{n,l}) (r_{n,k}\pm r_{n,l})
\Bigr)}\,,
\]
and similar for $n$ even, see \cite{gipst1}.

Including also the hitherto excluded case (iii), our final formula for the matrix element is
\[\fl
\frac{\langle \Phi_0|\sz{n}|\Phi_1\rangle
\langle \Phi_1|\sz{n}|\Phi_0\rangle}
{\langle \Phi_0|\Phi_0\rangle\langle \Phi_1|\Phi_1\rangle}
\;=\;(\lm_\pi^2-\lm_0^2)^{(D-\delta)/2} (\lm_0+\lm_\pi)^{\delta}
\prod_{{l<m}\atop{l,m \in \CD}}\!\!\lk
\frac{r_{l}+r_{m}}{r_{l}-r_{m}}\cdot\frac{\sin\,\frac{1}{2}(\qt_l-\qt_m)}{\sin\,\frac{1}{2}(\qt_l\,+\qt_m)}\rk
\times
\]
\bea\fl\lefteqn{\times\:\frac{\Lambda_n}{2^D \prod_{k\in\overline {\cal D}} (\dot + 2r_{k})}\cdot\frac
{\prod_{k{\rm \, odd},\, l{\rm \,even}}  \Bigl((\dot - r_{k}\dot +r_{l}) (\dot+r_{k}\dot- r_{l})\Bigr)}
{ \prod_{k<l, k,l{\rm \, odd}} \Bigl((\dot + r_{k} \dot +r_{l}) (\dot - r_{k}\dot -r_{l})\Bigr)
\prod_{k<l, k,l{\rm \, even}} \Bigl((\dot + r_{k}\dot+r_{l}) (\dot- r_{k}\dot- r_{l})\Bigr)},}\ny\\ && \label{prq}
\eea
where
\[\fl
\Lambda_n=\frac{\prod_{k\in \overline {\cal D}^\nul} (\lm_0\dot +r_k)}
{\prod_{k\in \overline {\cal D}^\one} (\lm_0\dot +r_k)\prod_{k\in\CD^\one} (\lm_0^2-r_k^2)}
\cdot
\frac{\prod_{k\in \overline {\cal D}^\one} (\lm_\pi\dot+r_k)}
{\prod_{k\in \overline {\cal D}^\nul} (\lm_\pi\dot+r_k)\prod_{k\in\CD^\nul} (\lm_\pi^2-r_k^2)}\,,\
\mbox{for odd $n$},
\]
\[\fl
\Lambda_n=\frac{\prod_{k\in \overline {\cal D}^\nul} (\lm_0\dot +r_k)(\lm_\pi\dot+r_k)}
{(\lm_0+\lm_\pi)\prod_{k\in \overline {\cal D}^\one} (\lm_0\dot +r_k)(\lm_\pi\dot+r_k)
\prod_{k\in\CD^\one} (\lm_0^2-r_k^2)(\lm_\pi^2-r_k^2)}\,,\qquad\hq
\mbox{for even $n$}.
\]
Here we used a superimposed dot:
$\:\dot \pm r_m\:$ as the short notation for $r_m$ if $m\in \CDC$, for $\pm r_m$ if $m \in \CD$ and for
$-r_m$ if $m\in \WD$, respectively. For composite sets of momentum levels $k$\\[1.5mm] we write
$\;\overline {\cal D}\:=\: \CDC\cup\WD$, $\;\;\overline {\cal D}^\nul\:=\:\CDC^\nul\cup \WD^\nul$,
$\;\;\overline {\cal D}^\one\:=\:\CDC^\one\cup \WD^\one$.

\subsection{Final result in terms of momenta}

Let $\{\qu_1$, $\qu_2,\ldots,$ $\qu_K\}$ and $\{\pu_1$, $\pu_2,\ldots$, $\pu_L\}$
be the sets of the momenta of the excitations presenting the states $|\Phi_0\rangle$
from the NS-sector and $|\Phi_1\rangle$ from the R-sector, respectively.
After some lengthy but straightforward
transformations of \r{prq} we obtain
\[\fl
\frac{\langle\, \Phi_0\,|\,\sz{n}\,|\,\Phi_1\,\rangle  \langle\,\Phi_1\,|\,\sz{n}\,|\,\Phi_0\,\rangle}
{\langle\, \Phi_0\,|\,\Phi_0\,\rangle \,\langle\,\Phi_1\,|\,\Phi_1\,\rangle}
\;=\; J \:(\lms_\pi+\lms_0)\, (\lms_\pi^2-\lms_0^2)^{(K+L-1)/2}\;\times\]
\be\fl \qquad\times\;
\prod_{k=1}^K
\frac{P^{\rm NS}_{\qu_k}\prod_{\qu\ne |\qu_k|}^{\frac{{\rm NS}}{2}} N_{\qu,\qu_k}}
{\prod_{\pu}^{\frac{{\rm R}}{2}} N_{\pu,\qu_k}}\cdot
\prod_{l=1}^L
\frac{P^{\rm R}_{\pu_l}\prod_{\pu\ne |\pu_l|}^{\frac{{\rm R}}{2}} N_{\pu,\pu_l}}
{\prod_{\qu}^{\frac{{\rm NS}}{2}} N_{\qu,\pu_l}}\cdot
\frac{\prod_{k=1}^K\prod_{l=1}^L M_{\qu_k,\pu_l}}
{ \prod_{k<k'}^K M_{\qu_k,\qu_{k'}}
\prod_{l<l'}^L M_{\pu_l,\pu_{l'}}}\,, \label{MEs}
\ee
where NS/2 (R/2) is the subset of quasi-momenta from NS (R) taking values in the segment $0<\qu<\pi$\,,~
NS/2 (R/2) containing $q_k$ with odd $k$ (even $k$):
\[ M_{\alpha,\beta} = \frac{\lms_\alpha+\lms_\beta}{\lms_\alpha-\lms_\beta}
\cdot\frac{\sin\frac{\alpha+\beta}{2}}{\sin\frac{\alpha-\beta}{2}}\,,\qquad
M_{\alpha,-\alpha} = \frac{\lms_\alpha^2\:
(\lms_0^2-\lms_\pi^2)}{(\lms_\pi^2-\lms_\alpha^2)(\lms_0^2-\lms_\alpha^2)}\,,
\]
\[
N_{\alpha,\beta} = \frac{\lms_\alpha+\lms_\beta}{\lms_\alpha-\lms_\beta},\hq {\cal J}\,=\:
\frac{{\prod_{\qu}}^{\!\!\!\frac{{\rm NS}}{2}}\:(\lms_0+\lms_\qu)}
  {{\prod_{\pu}}^{\!\!\!\frac{{\rm R}}{2}}\: (\lms_0+\lms_\pu)}\cdot
\frac{\prod_{\qu}^{\frac{{\rm NS}}{2}}\prod_{\pu}^{\frac{{\rm R}}{2}} (\lms_\qu+ \lms_\pu)^2 }
{ \prod_{\qu,\qu'}^{\frac{{\rm NS}}{2}}( \lms_\qu+ \lms_{\qu'})
\prod_{\pu,\pu'}^{\frac{{\rm R}}{2}} (\lms_\pu+ \lms_{\pu'})}\, .\]
For $n$ odd:\\[-11mm]
\[
P_{\qu}^{\rm NS}= \frac{\lms_\qu}{(\lms_\pi-\lms_\qu)(\lms_0+\lms_\qu)},\hx\qu\neq \pi,\hx
P_{\pu}^{\rm R} = \frac{\lms_\pu}{(\lms_\pi+\lms_\pu)(\lms_0-\lms_\pu)},\hx\pu\neq 0,\]
\[P_{0}^{\rm R} = P_{\pi}^{\rm NS} = \frac{1}{\lms_\pi+\lms_0},\hs
 J =\frac{{\prod_{\pu}}^{\!\!\!\frac{{\rm R}}{2}}\;(\lms_\pi+\lms_\pu) }
 {{\prod_{\qu}}^{\!\!\!\frac{{\rm NS}}{2}}\;(\lms_\pi+\lms_\qu)}\;\:{\cal J}
 ,\]
for $n$ even:\\[-11mm]
\[
P_{\qu}^{\rm NS} = \frac{\lms_\qu}{(\lms_\pi+\lms_\qu)(\lms_0+\lms_\qu)},\quad
P_{\pu}^{\rm R} = \frac{\lms_\pu}{(\lms_\pi-\lms_\pu)(\lms_0-\lms_\pu)},\hx\pu\neq 0,\,\pi,\]
\[ P_{0}^{\rm R} = -P_{\pi}^{\rm NS} = \frac{1}{\lms_\pi-\lms_0},\hs
 J =\frac{{\prod_{\pu}}^{\!\!\!\frac{{\rm NS}}{2}}\;(\lms_\pi+\lms_\qu) }
 {{\prod_{\qu}}^{\!\!\!\frac{{\rm R}}{2}}\;\:(\lms_\pi+\lms_\pu)}\;\:{\cal J}\,.\]

 \subsection{Bugrij--Lisovyy formula for the matrix elements}

In \cite{BL2} the following formula for the square of the matrix element of spin operator for the finite-size Ising model
was conjectured:
\[\fl
|\,{}_{\rm NS}\langle\, \qu_1, \qu_2,\ldots,\qu_K\,|\;\sigma^z_n\;|\,\pu_1,\pu_2,\ldots,\pu_L\,\rangle_{\rm R}|^2=
\]\[\fl
\hq=\;\xi\; \xi_T\; \prod_{k=1}^K \;\frac{\prod^{\rm NS}_{\qu\ne \qu_k} \sinh \frac{\g(\qu_k)+\g(q)}{2}}
{n \prod^{\rm R}_{\pu} \sinh \frac{\g(\qu_k)+\g(\pu)}{2}}\;\;
\prod_{l=1}^L \;\frac{\prod^{\rm R}_{\pu\ne \pu_l} \sinh \frac{\g(\pu_l)+\g(\pu)}{2}}
{n \prod^{\rm NS}_{\qu} \sinh \frac{\g(\pu_l)+\g(\qu)}{2}}\cdot
\left(\frac{t_y-t_y^{-1}}{t_x-t_x^{-1}}\right)^{\!\!(K-L)^2/2}\!\!\!\times
\]\be\fl\qquad\hq \times\;
\prod_{k<k'}^K \frac{\sin^2\frac{\qu_k-\qu_{k'}}{2}} {\sinh^2 \frac{\g(\qu_k)+\g(\qu_{k'})}{2}}
\;\;\prod_{l<l'}^L \frac{\sin^2\frac{\pu_l-\pu_{l'}}{2}} {\sinh^2 \frac{\g(\pu_l)+\g(\pu_{l'})}{2}}
\prod_{1\le k \le K \atop 1\le l \le L}
\frac {\sinh^2 \frac{\g(\qu_k)+\g(\pu_l)}{2}} {\sin^2\frac{\qu_k-\pu_l}{2}}\,.
\label{ME_BL}
\ee
In this formula the states are labelled by the momenta of the excitations.
The factors in front of the right hand side of \r{ME_BL} are defined by
\[\fl
\xi=((\sinh 2K_x \sinh 2K_y)^{-2}-1)^{1/4},\hx
\xi_T=\left(
\frac{\prod^{\rm NS}_{\qu} \prod^{\rm R}_{\pu} \sinh^2 \frac{\g(\qu)+\g(\pu)}{2}}
{\prod^{\rm NS}_{\qu,\qu'} \sinh \frac{\g(\qu)+\g(\qu')}{2}
\prod^{\rm R}_{\pu,\pu'} \sinh \frac{\g(\pu)+\g(\pu')}{2}} \right)^{1/4}\!\!,
\]
where $\gamma(\qu)$ is the energy of the excitation with quasi-momentum $\qu$:
\be\label{dr}\cosh \gamma(\qu)=\frac{(t_x+t_x^{-1})(t_y+t_y^{-1})}
{2(t_x^{-1}-t_x)}-\frac{t_y-t_y^{-1}}{t_x-t_x^{-1}}\,\cos \qu\,,
\ee
and $\:t_x\,=\,\tanh{K_x}$, $\;t_y\,=\,\tanh{K_y}$.

Formula \r{ME_BL} can be easily derived from \r{MEs} if one takes into account
the identification of parameters \r{ISI}. In particular we have
$t_x=a\,b$, $t_y=(a-b)/(a+b)$ and the relation
\be\label{gs}
e^{\gamma(\qu)}=\frac{a\, \lms_\qu\,+b}{a\, \lms_\qu\,-b}\,
\ee
between the energy $\,\gamma(\qu)\,$ of the excitation with quasi-momentum $\qu$ and
the corresponding zero $\lms_\qu$ of the $t(\lm)$-eigenvalue polynomial \r{tmgen}.
The following formulas give the correspondence between the different parts of \r{ME_BL} and \r{MEs}:
\[
\fl\frac{\xi\: \xi_T}{\sih{0}{\pi}}
\left(\frac{t_y-t_y^{-1}}{t_x-t_x^{-1}}\right)^{1/2}=J\,,\qquad
\frac{\sinh^2\frac{\g(\alpha)+\g(\beta)}{2}}{\sin^2\frac{\alpha-\beta}{2}}\;=\;
-\frac{t_y-t_y^{-1}}{t_x-t_x^{-1}}\;\: M_{\alpha,\beta}\,.
\]\[
\frac{\prod^{\rm NS}_{\qu\ne \qu_k} \sinh \frac{\g(\qu_k)+\g(\qu)}{2}}
{n \prod^{\rm R}_{\pu} \sinh \frac{\g(\qu_k)+\g(\pu)}{2}}=
\frac{\lms_0+\lms_\pi}{\sinh \frac{\g(0)+\g(\pi)}{2}}
\frac{P^{\rm NS}_{\qu_k}\prod_{\qu\ne |\qu_k|}^{\frac{{\rm NS}}{2}} N_{\qu,\qu_k}}
{\prod_{\pu}^{\frac{{\rm R}}{2}} N_{\pu,\qu_k}}\,,
\]\[
\frac{\prod^{\rm R}_{\pu\ne \pu_l} \sinh \frac{\g(\pu_l)+\g(\pu)}{2}}
{n \prod^{\rm NS}_{\qu} \sinh \frac{\g(\pu_j)+\g(\qu)}{2}}=
\frac{\lms_0+\lms_\pi}{\sinh \frac{\g(0)+\g(\pi)}{2}}
\frac{P^{\rm R}_{\pu_l}\prod_{\pu\ne |\pu_l|}^{\frac{{\rm R}}{2}} N_{\pu,\pu_l}}
{\prod_{\qu}^{\frac{{\rm NS}}{2}} N_{\qu,\pu_l}}\,,
\]
For more details, see \cite{gipst2}.

\section{Matrix elements for the diagonal-to-diagonal transfer-matrix
and for the quantum Ising chain in a transverse field}

In this section we derive the matrix elements of the spin operator between eigenvectors
of the diagonal-to-diagonal transfer-matrix for the Ising model on a square lattice (see Sect.~2.2).
In this case the parameters are given by \r{ISIdiag}.
As has been explained there, if we vary the parameters $a$ and $b$ in such a way
to have fixed $\:(a^2-b^2)/(1-a^2b^2)\,=\,1/{\sf k}',\;$ the eigenvectors
    (and therefore matrix elements) will not change.
So we fix $a=c={\sf k'}^{-1/2}$ and $\,b\,=d\,=0$.
Expanding the transfer-matrix \r{mm} with such parameters we get:
\[{\bf t}_n(\lm) ={\bf 1}-\frac{2\lm}{g}\:\widehat{\cal H}\:+\,\cdots\,, \qquad
 \widehat{\cal H}\,=\,-\frac{1}{2}\,\sum_{k=1}^n \:(\sz{k}\, \sz{k+1}\:+\,g\, \sx{k})\,,
\]
where $\widehat{\cal H}$
is the Hamiltonian of the periodic quantum Ising chain in a transverse field. {}From
\r{tmgen} we get the spectrum of this Hamiltonian:
\be\label{energy}
{\cal E}=-\frac{1}{2}\sum_\qu\pm\; {\ve(\qu)}
\ee
where the energies of the quasi-particle excitations are
\[\fl
\ve(\qu)=(1\:-2\,{\sf k'}\cos\qu+{\sf k'}^2)^{1/2}\;=\;\left(({\sf k'}-1)^2
\:+\, 4\,{\sf k'}\sin^2\frac{\qu}{2 }\right)^{1/2},\qquad \qu\ne 0,\:\pi\,,\]
\[\ve(0)\:=\:{\sf k'}-1,\quad\quad \ve(\pi)\:=\:{\sf k'}+1\,.\]
In \r{energy}, the sign $+/-$ in the front of  ${\ve(\qu)}$ corresponds
to the absence/presence of the excitation with the momentum $\qu$.
The NS-sector includes the states with an even number of excitations, the R-sector those with an odd number of excitations.
The momentum $\qu$ runs over the same set as in \r{tmgen}.
Since we have $a=c$ and $\:b=d$,~ the formula \r{MEs} with $\:\lms_\qu\:=\:{\sf k'}/\ve(\qu)$
for matrix elements for $\sz{n}$ can be applied. After some simplification we get the analogue
of \r{ME_BL}, now for the quantum Ising chain:
\[\fl
|\:{}_{\rm NS}\langle \qu_1,
\qu_2,\ldots,\qu_K\,|\,\sigma^z_m\,|\,\pu_1,\pu_2,\ldots,\pu_L\:\rangle_{\rm
R}|^2=  {\sf k'}^{\frac{(K-L)^2}{2}}\:\xi\; \xi_T \;\prod_{k=1}^K\: \frac{ \e^{\eta(\qu_k)}}{ n\: \ve(\qu_k)}\;\:\prod_{l=1}^L\:
\frac{ \e^{-\eta(\pu_l)}}{ n\: \ve(\pu_l)}\;\;
\times
\]
\be\fl \qquad\times\;
\prod_{k<k'}^{K}\left(\frac{2\sin \frac{\qu_k-\qu_{k'}}{2}}{\ve(\qu_k)+\ve(\qu_{k'})}\right)^2
\prod_{l<l'}^{L}\left(\frac{2\sin \frac{\pu_l-\pu_{l'}}{2}}{\ve(\pu_l)+\ve(\pu_{l'})}\right)^2
\prod_{k=1}^{K}\;\prod_{l=1}^{L}\:
\left(\frac{\ve(\pu_l)+\ve(\qu_k)}{2\sin \frac{\pu_l-\qu_k}{2}}\right)^2\,,\label{ME-IsCh}
\ee
where\\[-8mm]
\[
\xi\:=\: \left({\sf k'}^2-1\right)^{\frac 1 4}\,,\quad
\xi_T\:=\:\frac{\prod_{\qu}^ {\rm NS}\; \prod_{\pu}^ {\rm R}
(\ve(\qu)+\ve(\pu))^{\frac 1 2}}{\prod_{\qu,\qu'}^{\rm NS}\;(
\ve(\qu)+ \ve(\qu'))^{\frac 1 4}\;\prod_{\pu,\pu'}^{\rm R}\; (\ve(\pu)+
\ve(\pu'))^{\frac 1 4}}
\]
and\\[-8mm]
\[
\e^{\eta(\qu)}=\frac{ \prod_{\qu' }^{ \rm NS}\left(
\ve(\qu)+\ve(\qu')\right)}{\prod_{\pu}^{ \rm R} \left(\ve(\qu
)+\ve(\pu)\right)}\,.
\]
Formally, all these formulas are correct for the paramagnetic phase where ${\sf k'}>1$,
and for the ferromagnetic phase where $0\le {\sf k'}<1$. But for the case $0\le {\sf k'}<1\;$ it is natural
to redefine the energy of zero-momentum excitation as $\ve(0)=1-{\sf k'}$ to be positive.
{}From \r{energy}, this change of the sign of $\ve(0)$ in the ferromagnetic phase
leads to a formal change between absence-presence of zero-momentum excitation
in the labelling of eigenstates. Therefore
the number of the excitations in each sector (NS and R) becomes even.
Direct calculation shows that the change of the sign of $\ve(0)$ in \r{ME-IsCh}
can be absorbed to obtain formally the same formula \r{ME-IsCh},~ but with
new $\;\ve(0)$ and even $L$ (the number of the excitations in the R-sector) and
new $\:\xi\,=(1-{\sf k'}^2)^{1/4}$.

Formulas \r{ME_BL} and \r{ME-IsCh} allow to re-obtain
well-known formulas for the Ising model, e.g.
the spontaneous magnetization \cite{Onsager,Yang}.
Indeed, for the quantum Ising chain in the ferromagnetic phase
($0\le {\sf k'}<1$) and in the thermodynamic limit $n\to\infty$
(when the energies of $|\mbox{vac}\rangle_{\rm NS}$ and $|\mbox{vac}\rangle_{\rm R}$ coincide, giving
the degeneration of the ground state), we have $\xi_T\to 1$
and therefore the spontaneous magnetization
${}_{\rm NS}\langle\, \mbox{vac}\:|\:\sigma^z_m\:|\:\mbox{vac}\,\rangle_{\rm R}\;=\;\xi^{1/2}\;=\;(1-{\sf k'}^2)^{1/8}$.

\section{Conclusions}

We have shown that finite-size state vectors of the Ising (and generalized Ising) model can be obtained using the
method of Separation of Variables and solving explicitly Baxter equations. The Ising model is treated
as a special $N=2$ case of the $\ZN$-Baxter-Bazhanov-Stroganov $\tau^{(2)}$-model. Finite-size spin matrix elements between arbitrary
states are calculated by sandwiching the operators between the explicit form of the state vectors.
For the standard Ising case this gives a proof of the fully factorized formula for the form factors \r{ME_BL}
conjectured previously by Bugrij and Lisovyy.

We also extend this result to obtain a factorized formula for the
matrix elements of the finite-size Ising quantum chain in a transverse field. We show how specific local operators can
be expressed in terms of global elements of the monodromy matrix.
The truncated functional relation guaranteeing non-trivial
solutions of the Baxter equation is compared to Baxter's \cite{Baxter:book} Ising model functional relation.

\section*{Acknowledgements}

The authors are grateful to the organizers of the International Conference in memory of Alexei Zamolodchikov:
''Liouville field theory and Statistical models''
for their warm and friendly hospitality.
The authors wish to thank Yu.Tykhyy for his collaboration in \cite{gipst1,gipst2}.

G.v.G. and S.P. have been supported by the Heisenberg-Landau exchange program HLP-2008.
S.P. has also been supported in part by the RFBR
grant  08-01-00392 and the grant for the Support of Scientific Schools NSh-3036.2008.2.
The work of N.I. and V.S. has been supported in part by the grant
of the France-Ukrainian project 'Dnipro' No.M/17-2009, the grants of NAS of Ukraine, Special
Program of Basic Research and Grant No.10/07-N, the Russian-Ukrainian RFBR-FRSF grant.
G.v.G. thanks Yao-Zhong Zhang and the Department of Mathematics, University of Queensland, for kind hospitality.

\section*{References}
\bibliographystyle{amsplain}

\end{document}